\documentclass[conference]{IEEEtran}
\IEEEoverridecommandlockouts

\usepackage{cite}
\usepackage{amsmath,amssymb,amsfonts}
\usepackage{algorithmic}
\usepackage{graphicx}
\usepackage{textcomp}
\usepackage{xcolor}
\usepackage{multicol}

\usepackage{microtype}
\usepackage{subfigure}
\usepackage{booktabs}
\usepackage{algorithm}
\usepackage{flushend}
\usepackage{array}
\usepackage{float}
\usepackage{hyperref}

\usepackage{enumitem}
\def\BibTeX{{\rm B\kern-.05em{\sc i\kern-.025em b}\kern-.08em
    T\kern-.1667em\lower.7ex\hbox{E}\kern-.125emX}}

\usepackage{array}
\newcolumntype{P}[1]{>{\centering\arraybackslash}p{#1}}
\newcolumntype{M}[1]{>{\centering\arraybackslash}m{#1}}

\begin{document}

\title{CEDD-optimizer: Enabling Cost-Efficient \\Dataset Distillation on Geographically Distributed \\Edge Systems
}

\author{\IEEEauthorblockN{Dai Liu}
 \IEEEauthorblockA{\textit{Technical University of Munich}\\
 Garching, Germany \\
 dai.liu@tum.de}
 \and
 \IEEEauthorblockN{Eishi Arima}
 \IEEEauthorblockA{\textit{Technical University of Munich}\\
 Garching, Germany \\
 eishi.arima@tum.de}
 \and
 \IEEEauthorblockN{Martin Schulz}
 \IEEEauthorblockA{\textit{Technical University of Munich}\\
 Garching, Germany \\
 martin.w.j.schulz@tum.de}
}

\maketitle

\begin{abstract}
Centralized learning is a fundamental paradigm in modern AI, where data are typically collected from distributed edge devices and subsequently aggregated at a central host for model training. 
However, the overall training pipeline is often bottlenecked by the substantial communication overhead incurred during the data collection. 
Dataset Distillation (DD), benefiting from its remarkable compression ratio, has emerged as a leading dataset compression technique, making it particularly attractive for centralized learning on distributed data. 
However, while existing studies have demonstrated DD's overwhelming compression ratio, its cost efficiency in non-uniform edge environments has been largely overlooked. 
As edge devices are often distributed geographically in the real world, their energy and data transfer prices are often non-uniform. 
At the same time, several key hyperparameters in DD (e.g., the target compression ratio and the number of distillation steps) affect the energy and data transfer overhead considerably as well as the training quality (or the test accuracy of downstream training tasks), requiring careful tuning both locally and globally. 
In this work, we propose \textit{\underbar{C}ost-\underbar{E}fficient \underbar{D}ataset \underbar{D}istillation optimizer (CEDD-optimizer)}, a hyperparameter tuning framework for cost-efficient distributed DD. 
Our framework aims at minimizing the total cost under a constraint for training quality by optimizing the hyperparameter settings across edge devices while being aware of the environmental non-uniformity. 
Our framework relies on two key modules: \textit{CEDD-calibrator} and \textit{CEDD-solver}. 
The CEDD-calibrator identifies parameters in our energy and training quality modeling --- the former is detected by an offline calibration, while the latter is estimated online during our three-step tuning scheme. 
Building upon these models, the CEDD-solver deals with the cost minimization problem to steer and improve the distributed DD workflow. 
Our thorough experiments across various image datasets show that our approach achieves up to a 20.8x improvement over the baseline DD method under the same quality constraint.
\end{abstract}


\section{Introduction}
Deep learning~\cite{deep-learning} has been widely adopted across a broad range of applications, all of which rely on training deep neural networks with large amounts of data. 
A common training paradigm is centralized learning~\cite{centralized-learning}, where a training dataset, typically collected from edge devices distributed across different locations, is aggregated and used to train a model at a central host. 
Real-world applications rely on this paradigm, e.g., for edge-cloud defect detection in manufacturing~\cite{usecase1}, privacy-preserving federated histopathology across hospitals~\cite{usecase3}, distributed crop monitoring in agriculture~\cite{usecase2, usecase4}, managing heterogeneous edge devices via centralized slimmable networks~\cite{usecase5}.
However, the training pipelines are often costly and constrained by communication overhead when transferring large datasets from edge devices to the host. 

Traditional lossy and lossless compression techniques~\cite{compression1,compression2,compression3} do not effectively resolve the communication burden. 
On one hand, lossless compression techniques alone do not mitigate the data transfer bottleneck for deep learning tasks due to their limited compression ratios. 
On the other hand, the deep neural network training process  relies on subtle and fine-grained details in the data for effective learning, whereas lossy compression methods, which achieve higher compression ratios, inevitably sacrifice these details, resulting in poor training quality.


\textbf{Dataset Distillation (DD)}~\cite{dd-original,dd-dc,dd-fed2} 
has emerged as a promising alternative to traditional dataset compression techniques. 
Unlike traditional training tasks, 
DD utilizes network gradients to train a smaller synthetic dataset. 
Compared to traditional compression methods, DD offers the following major advantages: (i) very high compression ratio~\cite{dd-original,dd-dc,dd-cafe,dd-att,dd-fed2,dd-datm,dd-am}; (ii) marginal accuracy loss~\cite{dd-datm,dd-att}; (iii) data anonymization properties by synthesizing data from the original dataset~\cite{dd-privacy,dd-privacy1,dd-privacy4}; and (iv) no decompression required, thus a significant overhead reduction for downstream training tasks including continual learning~\cite{dd-continue1,dd-continue2,dd-continue3,dd-continue4,dd-continue5,dd-continue6}. 
These properties collectively address the aforementioned obstacles, making DD an excellent fit for efficient centralized learning scenarios.

Despite its substantial benefits, it relies on a costly hyperparameter tuning step, which is a major challenge when deploying DD in distributed edge environments. This, however, has been largely overlooked in the literature. 
In particular, two key hyperparameters significantly affect both test accuracy and operational cost: the target compression ratio and the distillation loop iteration count. 
However, in existing DD methods~\cite{dd-fed2,dd-fed-4,dd-fed-3}, these parameters are manually tuned without accounting for real-world environments, where systems are typically non-uniform, particularly in terms of data transfer and electricity costs, depending on the geographical locations of edges. 
Therefore, merely deploying conventional DD methods results in a suboptimal hyperparameter setup (e.g., assigning a high DD compute load to an energy-intensive region) when facing a key optimization challenge: minimizing the total operational cost while satisfying a requirement for model quality. 






\begin{figure}[t]
\begin{center}
  \includegraphics[clip, trim=0cm 0cm 0cm 0cm,width=1\linewidth]{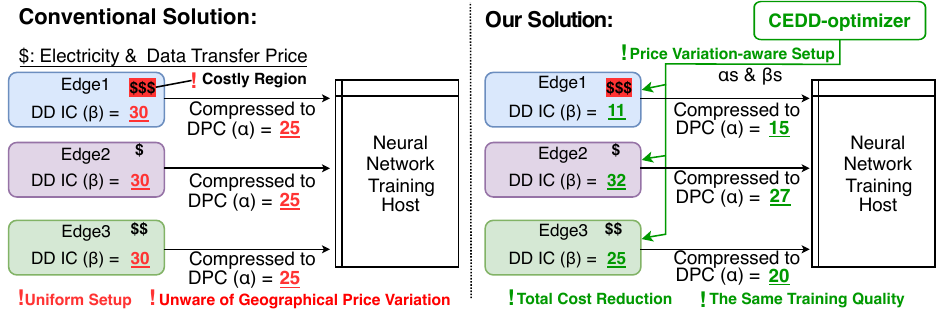}
  \caption{The Basic Concept of Our Work: Price Variation-aware Tuning for Data Per Class (DPC, $\alpha$) and Iteration Count (IC, $\beta$)}
  \label{fig:solution-overview}
\end{center}
\end{figure}

In this work, we focus on this significant optimization challenge and present \textit{\underbar{C}ost-\underbar{E}fficient \underbar{D}ataset \underbar{D}istillation optimizer (CEDD-optimizer)}, a novel hyperparameter tuning framework, with a particular focus on distributed DD with non-uniform device setups. 
The basic concept is illustrated in Fig.~\ref{fig:solution-overview}. 
Our work targets a centralized learning pipeline in which each edge device applies DD to its local data to produce a compact, distilled dataset. 
These distilled datasets are then aggregated at a central training host 
to train a deep neural network. 
Our CEDD-optimizer optimizes the two key DD hyperparameters for each edge node separately: (i) distilled data per class (DPC, $\alpha$), which determines the volume of transferred data, and (ii) iteration count (IC, $\beta$), which governs local computation and energy consumption. 

To this end, it coordinates two key components: an offline CEDD-calibrator and an online CEDD-solver. 
The former efficiently identifies the unknowns of two predictive models used to estimate energy consumption (offline) and test accuracy (online). 
The CEDD-solver then employs these predictive models to solve the optimization problem that minimizes total system cost (including communication, energy, and storage) for a given target test accuracy for downstream deep learning tasks. 
By selecting the optimal pair $(\alpha, \beta)$ for each edge device separately, our CEDD-optimizer provides tailored configurations that align the quality of the aggregated distilled dataset with the system cost-efficiency across non-uniform, geographically distributed deployments. 

The following are the major contributions of this paper: 
\begin{itemize}[noitemsep, labelindent=0pt, leftmargin=*]
\item \textbf{Identification of the new challenge: }
To the best of our knowledge, this is the first work to identify the challenge of hyperparameter tuning in DD under non-uniform, geographically distributed edge environments, accounting for the total economic cost arising from heterogeneous networks and energy prices across geographical locations. 

\item \textbf{Formulation and modeling: }
We formulate the hyperparameter tuning challenge as a formal optimization problem in a concrete mathematical format. 
We construct simple predictive models to estimate the test accuracy of deep learning tasks trained with the distilled dataset, the energy consumption of DD on the edges, and the volume of the distilled dataset as functions of DD's hyperparameters. 

\item \textbf{The CEDD-optimizer framework: }
We offer a cost-aware hyperparameter tuning framework for distributed DD that coordinates the CEDD-calibrator and the CEDD-solver. 
The former efficiently identifies the model coefficients offline or online, while the latter solves the optimization problem using the models. 

\item \textbf{Thorough evaluation: }
We thoroughly evaluate and validate the proposed models and optimizer in diverse configurations, using four different edge devices, five benchmark datasets, 
and a wide range of geographically distributed price data. 
Our results demonstrate that CEDD-optimizer consistently achieves significant cost reductions while maintaining competitive accuracy, and we further analyze its scalability and trade-offs under varying system conditions.
Find our code in \url{https://github.com/NiaLiu/CEDD-optimizer.git}.
\end{itemize}


\section{Background and Related Work}
\subsection{Dataset Distillation}\label{sub:dd-literature}
Dataset Distillation (DD) was first proposed by Wang et al. in 2018~\cite{dd-original} as an efficient data compression method specialized for model training in deep learning tasks. 
Various follow-up studies~\cite{dd-dc,dd-dm,dd-cafe,dd-mtt,dd-att} have proven that DD significantly outperforms conventional lossy or lossless dataset compression methods, including the Coreset Selection~\cite{dd-dc,dd-cafe,dd-mtt} and others~\cite{cs-herding,cs-herding1,cs-forgetting,cs-forgetting1,cs1}. 
Recent DD algorithms can be split into iterative methods and non-iterative methods. The iterative methods are considered variants of DC (Dataset Condensation)~\cite{dd-dc}. It typically relies on iterative bi-level optimization, such as aligning feature distributions~\cite{dd-dm}, addressing classes' miss-alignment~\cite{dd-idm,dd-dm}, capturing classes' differences~\cite{dd-dcc}, aligning layer-wise features~\cite{dd-cafe}, embedding differentiable siamese augmentation~\cite{dd-dsa}, applying a kernel-based meta-learning framework~\cite{dd-kip}, and adopting soft labels~\cite{dd-sl, dd-sl1, dd-tasla}, etc.~\cite{guo2024lossless,lee2024selmatch,zhong2025stable,chen2025curriculum}. Further, non-iterative methods are often one shot~\cite{wang2023dim,su2024d4m,renuga2024medsynth,latentvideo2025,tabular2025}.

In this paper, we focus on iterative methods, which are generally costly but provide good results. 
For generality, we choose the original DC~\cite{dd-dc} as our baseline. 
Our approach is not specific to DC and is generally applicable to other iterative DD methods, as we target key parameters commonly available in any DD method.
\textit{Overall, our approach is the first to introduce a geographical location-based hyperparameter tuning methodology in DD to minimize the total cost with marginal accuracy loss for edge deployments.}



\subsection{Edge Computing}\label{sub:etcc-literature}
The concept of edge computing has gained momentum ever since it was proposed in the 2000s~\cite{edge-history}, and various studies have focused on big data-driven AI applications in edge computing.   
Several recent studies target cost management for edge computing or the edge-to-cloud continuum~\cite{long2020game, huang2023cost, li2023cost, rac2024cost}. 
Lu~et~al.~\cite{edge-cloud-lu} propose a hybrid method that combines lossy and lossless compression for edge computing, and Wu~et~al.~\cite{edge-cloud-wu}  propose a CNN-based encoder and decoder. 
\textit{Our work introduces DD as a promising alternative.}
There are several tools for service deployment in edge computing. 
For example, Rosendo~et~al.~\cite{edge-cloud-rosendo} provide E2Clab. 
They later extend the framework to support an optimization functionality~\cite{edge-cloud-rosendo2} and enable efficient provenance capture in IoT/Edge~\cite{edge-cloud-rosendo3} environments. 
Some studies spot on environmental constraints in the edge-to-cloud continuum, including (1) network bandwidth constraint and network slicing optimization~\cite{edge-cloud-habeeb} and (2) energy constraint for battery-powered or energy-harvesting devices~\cite{edge-cloud-battery, energy-harvesting, energy-harvesting2}. Others target heterogeneous edge computing environments~\cite{edge-cloud-hetero,edge-cloud-hetero2}.

It is worth noting that our work is orthogonal to decentralized learning methods, such as Federated Learning (FL)~\cite{Fedavg,FedNova}. 
While DD can be integrated into FL frameworks to enhance communication efficiency and data privacy~\cite{dd-fed0,dd-fed-early1,dd-fed2,dd-fed-3,dd-fed-4,dd-fed-early2}, this integration lies beyond the scope of this study. 
\textit{
In this paper, we focus instead on improving the efficiency of centralized learning pipelines, where DD serves as the core mechanism for data compression and quality/cost optimization.
}

\section{Introducing Dataset Distillation to Edge Systems}\label{motivation}


\begin{figure}[t]
\centering
  \includegraphics[width=\linewidth]{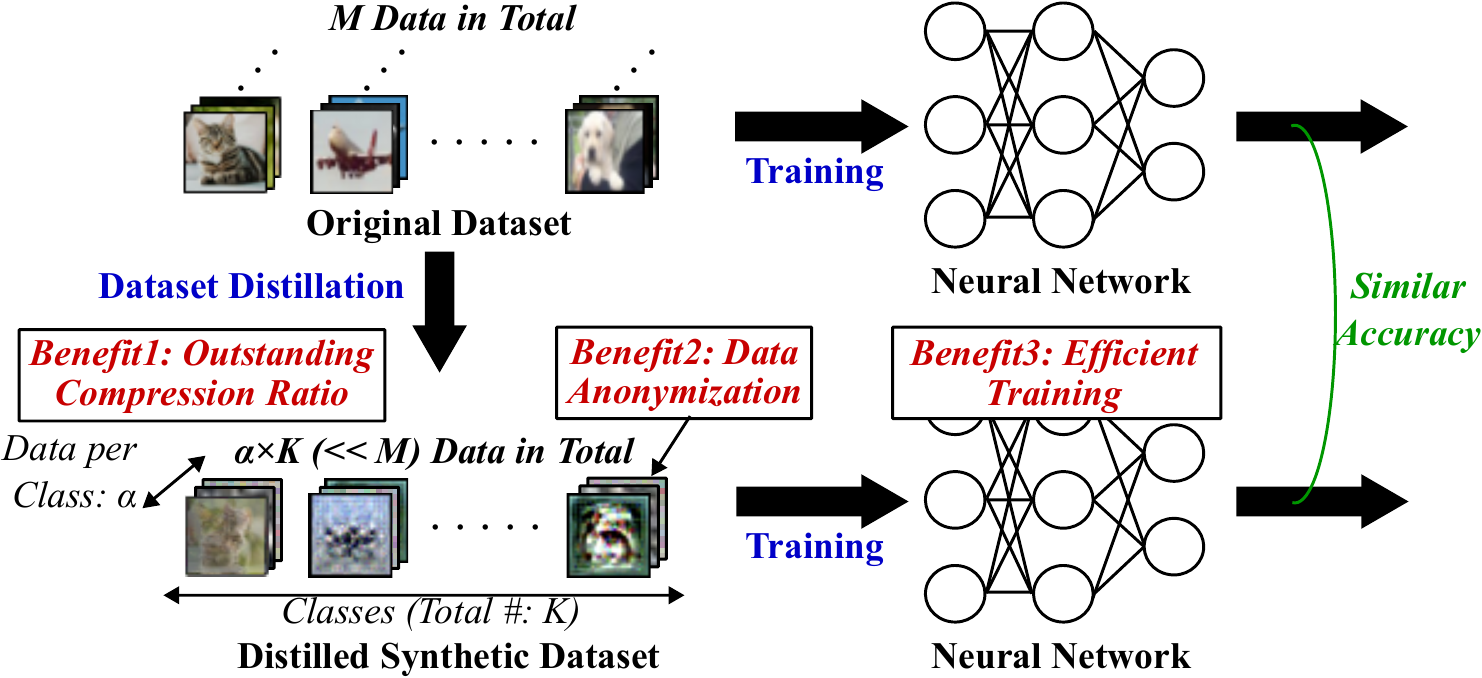}
  \caption{Concept and Benefits of DD: Replacing the Original Dataset with a Synthetic One for Efficient Training}
  \label{fig:dd-concept}
\end{figure}

\begin{figure}[t]
\centering
    \includegraphics[width=\linewidth]{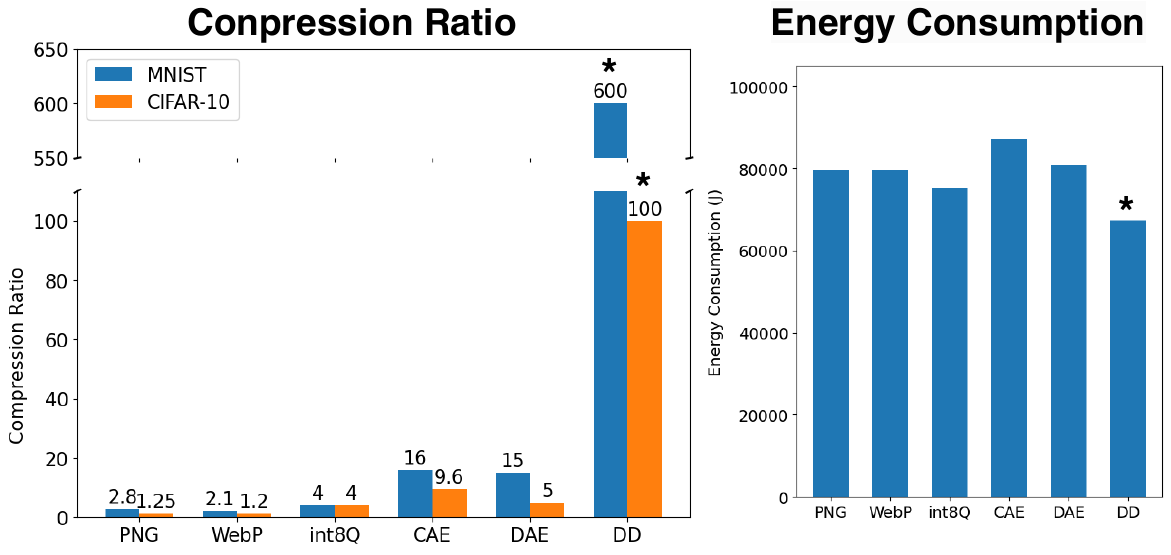}
    \caption{Comparisons of Compression Ratio and Energy among Various Compression Methods}
    \label{fig:DD-prove3}
\end{figure}

\subsection{Motivation: High Compression Ratio of DD}\label{sec:dd-concept}

Fig.~\ref{fig:dd-concept} depicts the overall concept of DD. 
DD converts an original dataset ($M$ data in total) into a much smaller number of \textit{synthetic} distilled data. Subsequently, the distilled dataset is used to train a given neural network.
As the synthetic distilled dataset conceals the training information of the original dataset, the neural network trained on the distilled dataset can achieve inference accuracy similar to that of the original dataset. 
DD has been applied to classification tasks in the literature, and the number of data per class (or DPC) is a key hyperparameter that determines the overall data compression ratio. 
Let $M$, $K$, and $\alpha$ be the total size of the original dataset, the number of classes in the dataset, and the DPC in the distilled dataset, respectively. 
Then, the compression ratio is simply denoted as $M/(\alpha\cdot K)$. 
Therefore, the compression ratio in DD is fully controllable by the hyperparameter $\alpha$.

Fig.~\ref{fig:DD-prove3} compares the compression ratio and energy consumption among DD and other existing lossless or lossy compression mechanisms: PNG~\cite{png}, WebP, int8Q (an int8 quantization)~\cite{int8q}, CAE (ConvAE, a classic convolutional autoencoder)~\cite{autoencoder}, and DAE (DenseAE, a fully connected autoencoder)~\cite{dense-autoencoder}. 
As shown in the left graph of Fig.~\ref{fig:DD-prove3}, DD achieves an outstanding compression ratio, outperforming the others by \textbf{orders of magnitude improvement}. 
In this evaluation, we used a simple three-layer convolutional neural network as the baseline model for both DD and CAE. 
We set the test accuracy thresholds for training tasks to 97\% and 60\% for the MNIST~\cite{mnist} and CIFAR-10~\cite{cifar10} datasets, respectively. 
As for the energy consumption, although DD requires a noticeable energy overhead for its compression procedure, which is also controllable by tuning hyperparameters, it can significantly reduce the overhead for downstream model training using the distilled dataset. 
As a consequence, the total energy cost of DD is the smallest in this experiment. 
In this energy evaluation, we use the same platform (NVIDIA Jetson Xavier NX) for all computations, which will be described later in Section~\ref{evaluation}.

\begin{figure}[t]
  \begin{center}
  \includegraphics[width=\linewidth]{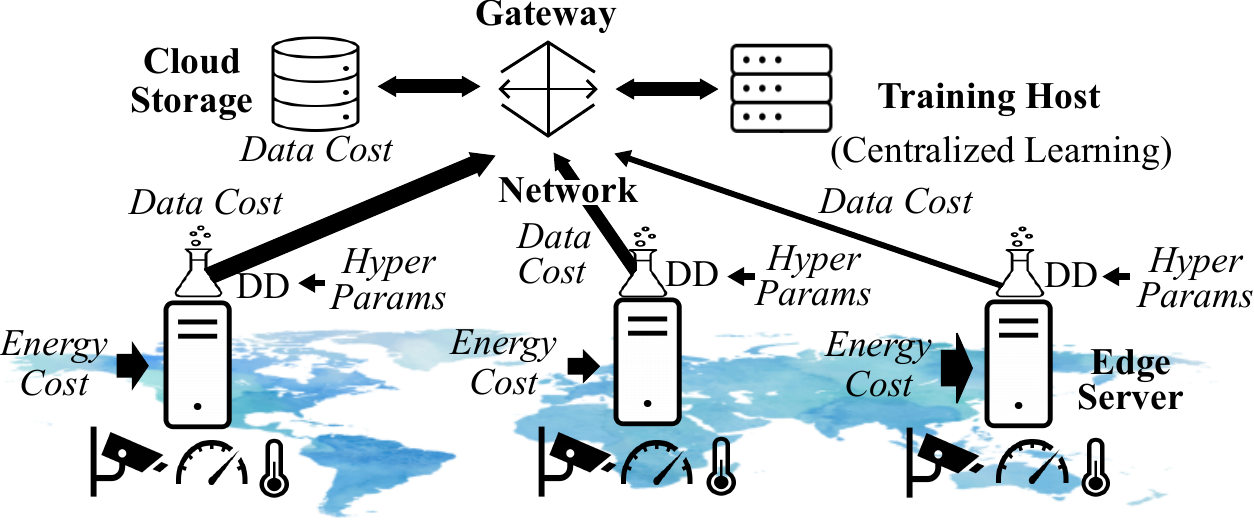}
  \caption{Our Target System Architecture and DD Scenario with Data / Energy Cost Management via Hyperparameter Tuning}
  \label{fig:overview}
  \end{center}
\end{figure}

\subsection{Application Scenario}\label{scenario}

Fig.~\ref{fig:overview} illustrates the scenario we assume in this paper.
We target the following computing continuum: (1) data are generated on end devices geographically distributed across the world; (2) each edge server receives the data from local end devices and applies DD to the collected dataset; (3) the distilled dataset is aggregated and stored on the cloud storage; (4) the training host utilizes the distilled dataset to train a neural network model; and (5) the trained model is stored on the cloud storage. 
Note, this scenario is based on centralized learning, commonly used in a wide variety of services~\cite{fl-vs-cl}. 
We also account for storage costs in our experiments to cover the potential of continual learning~\cite{dd-continue1,dd-continue2,dd-continue3,dd-continue4,dd-continue5,dd-continue6}.
In the target scenario, \textit{we use CEDD-optimizer to tune the hyperparameters of DD distributed across edges to manage the economic cost encompassing data transfer/storage and electricity for DD, which are highly dependent on the geographical locations of edge devices.}

\subsection{Leveraging Hyperparameters in DD}\label{trade-off}
\begin{figure}[h]
\centering
  \includegraphics[width=\linewidth]{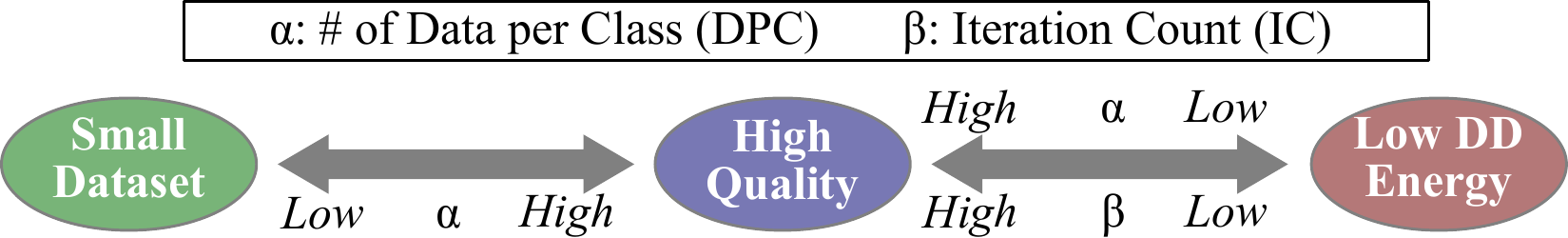}
  \caption{Trade-off Relationships in DD among Dataset Size, Quality, and Energy Overhead}
  \label{fig:trade-off}
\end{figure}

We leverage the key hyperparameters that govern the quality of the distilled dataset and the DD overhead. 
The objective is to minimize the cost of data transfer/storage and energy consumption at the edges while meeting a target quality requirement. 
As a \textbf{quality metric}, we use the \textbf{test accuracy} of downstream deep learning tasks trained with the distilled dataset, as it reflects the quality of the dataset and matters for the users as well. 
To this end, we deal track and adjust the two following parameters across all edge systems: (1) the number of data per class (or DPC) denoted as $\alpha$ and (2) the upper bound of iteration count for the outermost loop of DD (or IC) represented as $\beta$. 
Fig.~\ref{fig:trade-off} illustrates the trade-off relationships governed by these two types of hyperparameters. 
Both parameters affect the test accuracy of downstream deep learning tasks and the energy (or computational) overhead to perform DD on the edge, whereas the dataset size is affected only by the DPC (see also Section~\ref{sec:dd-concept}). 

On one hand, DD is typically implemented as an iterative solver, and thus the execution time and energy consumption are linear functions of the IC (or $\beta$) that determines the count of outermost loop. 
On the other hand, the computational and energy overheads of DD per loop are determined by the number of data per class (DPC or $\alpha$).
In particular, they can be modeled by a quadratic function of DPC $\alpha$.
This is because the DD main loop processes $\alpha$ images x $\alpha$ times per iteration, resulting in a computational complexity of $\mathcal{O}(\alpha^2)$ per iteration. We model and verify this later in Section~\ref{energy-formulation}~and~\ref{subsub:validate energy prediction}.

\begin{figure}[t]
\begin{center}
  \includegraphics[width=\linewidth]{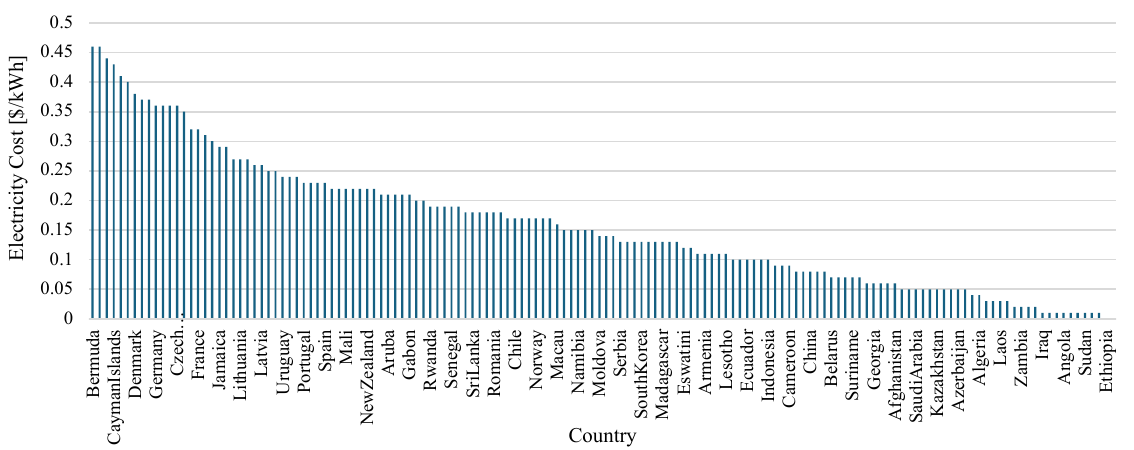}
  \caption{Electricity Cost by Country in 2024 including 144 Countries (Remade from~\cite{electricity})}
  \label{fig:electricity-cost}
\end{center}
\vspace{-5pt}
\begin{center}
  \includegraphics[width=\linewidth]{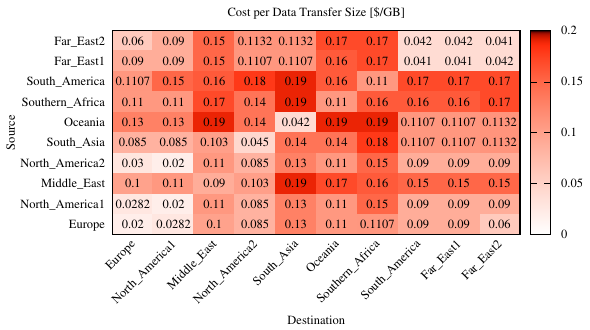}
  \caption{Data Transfer Cost (or SiteLink Rate) by Location in AWS Direct Connect (Remade from~\cite{aws-price})}
  \label{fig:sitelink_plot}
\end{center}
\end{figure}

\subsection{Impact of Geographical Locations on Pricing}\label{sec:pricing}
We target the DD scenario illustrated in Fig.~\ref{fig:overview} where end devices and edge devices are geographically distributed at a global scale while being aware of both the total operating cost and the test accuracy of deep learning tasks. 
\textit{This global deployment results in variations of electricity cost for edge devices and data transfer pricing per size. }
Several recent studies also address cost management in edge computing or the edge-to-cloud continuum; however, they do not consider DD or the price variations of edge energy and data transfer simultaneously~\cite{long2020game, huang2023cost, li2023cost, rac2024cost}. For example, the study by Huang~et~al.~\cite{huang2023cost} considers the costs of energy and communication, but does not account for their variations. 
\textit{To the best of our knowledge, our work is the first to target the combination of the geographical price variations and the application of DD to edge devices with hyperparameter tuning. }

Fig.~\ref{fig:electricity-cost} lists the electricity price [\$/kWh] by country in 2024, encompassing 144 countries~\cite{electricity}, while Fig.~\ref{fig:sitelink_plot} demonstrates the communication cost per combination of source and destination locations [\$/GB] in a global private network infrastructure service, in this case AWS direct connect~\cite{sitelink,aws-price} as an example. 
In Fig.~\ref{fig:electricity-cost}, the X-axis lists countries sorted by cost in descending order, whereas the Y-axis represents the electricity cost. 
As shown, the variation is significant by orders of magnitude. 
As for communication costs, we show the operating cost per data transfer when using a global private network provided by AWS~\cite{sitelink} as an example. 
It enables both edge-to-edge and edge-to-cloud communications in a  direct and private fashion, while charging in a pay-as-you-go policy~\cite{sitelink,aws-price}. 
As shown in Fig.~\ref{fig:sitelink_plot}, the communication cost varies by an order of magnitude depending on the geographical locations of edge devices. 

\section{Formulation and Modeling}\label{problem}


\subsection{Problem Formulation}\label{formulation}
\begin{figure}[t]
\begin{center}
  \includegraphics[width=\linewidth]{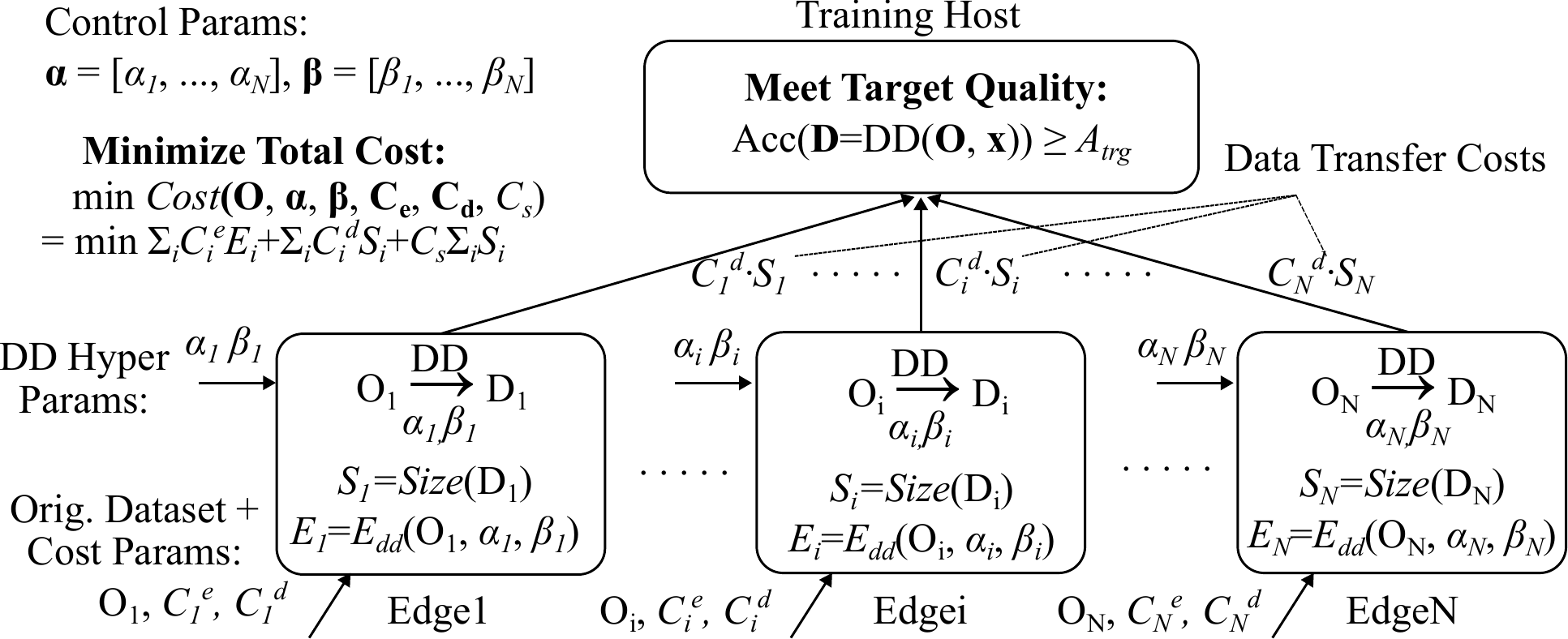}
  \caption{Problem Overview: Minimizing the Total Cost ($Cost$) for a Target Test Accuracy ($Acc$) by Leveraging Hyperparameters across Edges ($\boldsymbol{\alpha}=[\alpha_1, \cdots, \alpha_N]$, $\boldsymbol{\beta}=[\beta_1, \cdots, \beta_N]$)}
  \label{fig:problem}
\end{center}
\end{figure}

Fig.~\ref{fig:problem} illustrates the overall problem we address with this study. 
We target the system that comprises $N$ edge nodes and one training host that performs the downstream learning task using the collected dataset from the edge nodes. 
On the $i$th edge node, the hyperparameters $\alpha_i$ (or DPC) and $\beta_i$ (or IC) need to be set accordingly, depending on given inputs or requirements. 
The following parameters are associated with the $i$th node: $\mathrm{O_i}$, $C_i^e$, and $C_i^d$. 
Here, $\mathrm{O_i}$ is the original dataset collected at the node, whereas $C_i^e$ or $C_i^d$ are the cost parameters that represent the price per energy [\$/J] or the price per transferred data volume [\$/B], respectively. 
As mentioned in Section~\ref{sec:pricing}, these cost parameters can vary significantly by location. 
Once the hyperparameters $\alpha_i$ and $\beta_i$ are set, the DD is applied to $\mathrm{O_i}$ to generate a distilled dataset $\mathrm{D_i}$. 
The aggregated distilled datasets $\mathbf{D}=[\mathrm{D_1}, \mathrm{D_2}, \cdots, \mathrm{D_N}]$ are used in the downstream training task. 
Thus, its test accuracy $Acc()$ is a function of the aggregated distilled datasets $\mathbf{D}$. 
In this study, we deal with the following optimization:  minimizing the total cost $Cost()$ for a given targeted accuracy $Acc() \geq A_{trg}$. 
To this end, the tuning procedure determines $\boldsymbol{\alpha}=[\alpha_1, \cdots, \alpha_N]$, $\boldsymbol{\beta}=[\beta_1, \cdots, \beta_N]$, tailored for the set of original datasets $\mathbf{O}=[\mathrm{O_1}, \mathrm{O_2}, \cdots, \mathrm{O_N}]$ and given price parameters including (1) the edge energy costs per Joule $\mathbf{C_e}=[C^e_1, C^e_2,\cdots,C^e_N]$, (2) the data transfer costs per Byte between the training host and edge devices $\mathbf{C_d}=[C^d_1, C^d_2,\cdots,C^d_N]$, and (3) the cloud storage cost per Byte $C_s$. 

The optimization problem is formulated as follows: 
\begin{eqnarray}
&In:& \mathbf{O}, \mathbf{C_e}, \mathbf{C_d}, C_{s}\nonumber\\ 
&Out:& \boldsymbol{\alpha}=[\alpha_1, \cdots, \alpha_N], \boldsymbol{\beta}=[\beta_1, \cdots, \beta_N] \nonumber\\
&\min& Cost(\mathbf{O}, \boldsymbol{\alpha}, \boldsymbol{\beta}, \mathbf{C_e}, \mathbf{C_d},  C_{s})\\
& s.t. & Acc(\mathbf{D} = \mathbf{DD}(\mathbf{O}, \boldsymbol{\alpha}, \boldsymbol{\beta})) \geq A_{trg}\\
& & \alpha_{min} \leq \alpha_{i} \leq \alpha_{max} = M/(N\cdot K\cdot R_{trg})\label{eq:alpha-range}\\& & \beta_{min} \leq \beta_{i} \leq \beta_{max}~~(1 \leq {}^\forall i \leq N)\label{eq:hp-range}
\end{eqnarray}

The objective is to minimize $Cost()$ that encompasses the costs of data transfers, edge energy, and cloud storage, which is a function of $\mathbf{O}$, $\boldsymbol{\alpha}$, $\boldsymbol{\beta}$, $\mathbf{C_e}$, $\mathbf{C_d}$, and $C_{s}$. 
Eq.~(\ref{eq:alpha-range}) and Eq.~(\ref{eq:hp-range}) show the ranges of DPC $\alpha_i$ and IC $\beta_i$, respectively, and the ranges of DPC and IC are equally set to all nodes. 
Note that $\alpha_{max}$ is set in accordance with a given lower limit compression ratio per node $R_{trg}$ (see also Section~\ref{sec:dd-concept}), in order to explicitly control the amount of data transfer traffic. 
The parameters and functions used in our formulations are summarized in Table~\ref{table:refs}. 

\begin{table}[b]
\begin{center}
\caption{Notations of Variables/Functions}
\label{table:refs}
{
\scriptsize
\begin{tabular}{|M{0.09\linewidth}||M{0.79\linewidth}|}
    \hline
    Variable & Remarks \\
    \hline\hline
    $N$, $M$, $K$, $V$ & \# of edge devices, the total \# of original data across edges, \# of classes in the classification task, the size per data  \\
    \hline
    \hline

    $\alpha_i$ & DPC (Data Per Class) on the $i$th edge ($\alpha_{min}\leq\alpha_i\leq\alpha_{max}$) \\
    \hline
    $\beta_i$ & IC (Iteration Count) on the $i$th edge ($\beta_{min}\leq\beta_i\leq\beta_{max}$)\\
    \hline
    $\boldsymbol{\alpha}$, $\boldsymbol{\beta}$ & The vector of $\alpha_i,\beta_i$: $\boldsymbol{\alpha} = [ \alpha_1, \cdots, \alpha_N], \boldsymbol{\beta} = [\beta_1,\cdots,\beta_N]$ \\  
    \hline\hline
    $\mathbf{O}$ & The list of original datasets: $\mathbf{O} = [ \mathrm{O_1}, \mathrm{O_2}, \cdots, \mathrm{O_N}]$ ($\mathrm{O_i}$: The original dataset at the $i$th edge)\\
    \hline
    $\mathbf{D}$ & The list of distilled datasets: $\mathbf{D} = [ \mathrm{D_1}, \mathrm{D_2}, \cdots, \mathrm{D_N}]$ ($\mathrm{D_i}$: The distilled dataset at the $i$th edge)\\
    \hline
    \hline
    $A_{trg}$ & The parameter to setup the target accuracy \\
    \hline
    $R_{trg}$ & The parameter to setup the lower limit compression ratio per node \\
    \hline
    $\mathbf{C_e}$ & The vector of edge energy prices [\$/J]: $\mathbf{C_e} = [ C^e_1, \cdots, C^e_N]$ \\
    \hline
    $\mathbf{C_d}$ & The vector of data transfer prices [\$/B]: $\mathbf{C_d} = [ C^d_1, \cdots, C^d_N]$ \\
    \hline
    $C_{s}$ & The cloud storage cost per size (e.g., \$ 0.023 per GB~\cite{storage-price})\\
    \hline
    $\mathbf{E}$ & The vector of DD energy overhead: $\mathbf{E} = [ E_1, \cdots, E_N]$ \\
    \hline
    $\mathbf{S}$ & The vector of data size to transfer: $\mathbf{S} = [ S_1, \cdots, S_N]$ \\ 
    \hline\hline


\end{tabular}
\begin{tabular}{|M{0.12\linewidth}||M{0.76\linewidth}|}
    \hline
    Function & Remarks \\
    \hline\hline
    $Acc()$ & The test accuracy for the aggregated distilled dataset $\mathbf{D}=[\mathrm{D_1},\cdots, \mathrm{D_N}]$ \\\hline
    $Cost()$ & The total cost as a function of $\mathbf{O}, \boldsymbol{\alpha},\boldsymbol{\beta}, \mathbf{C_e}, \mathbf{C_d},  C_{s}$ \\\hline
    $E_{dd}()$ & DD's energy overhead ($\mathrm{O_i}\rightarrow\mathrm{D_i}$) as a function of $O_i$, $\alpha_i$, $\beta_i$ \\
    \hline
    $DD()$ or $\mathbf{DD}()$ & Dataset distillation ($\mathrm{O_i}\rightarrow\mathrm{D_i}$) or ($\mathbf{O}\rightarrow\mathbf{D}$) as a function of ($\mathrm{O_i}$ $\alpha_i$, $\beta_i$) or ($\mathbf{O}, \boldsymbol{\alpha}, \boldsymbol{\beta}$) \\\hline
    $Size()$& The size of the given dataset $D_i$ or set of dataset $\mathbf{D}$\\\hline
\end{tabular}
}
\end{center}
\end{table}

\subsection{Cost and Accuracy Modeling}\label{modeling}
In order to optimize the hyperparameter setup ($\boldsymbol{\alpha}, \boldsymbol{\beta}$) by solving the problem formulated in Section~\ref{formulation}, both the cost and accuracy functions ($Cost()$ and $Acc()$) need to be known. 
In this section, we present our predictive modeling.

\subsubsection{\bf Cost Modeling}\label{energy-formulation}
As presented in Fig.~\ref{fig:problem}, the total cost is broken down into (1) energy cost, (2) data transfer cost, and (3) storage cost in this study. 
We model the cost function by using three dot-product terms as follows: 
\begin{eqnarray}
&~~&Cost(\mathbf{O}, \boldsymbol{\alpha}, \boldsymbol{\beta}, \mathbf{C_e}, \mathbf{C_d},  C_{s})\nonumber\\
&=& \mathbf{C_e}\cdot\mathbf{E} + \mathbf{C_d}\cdot\mathbf{S} + C_{s}\mathbf{1}\cdot\mathbf{S}\nonumber\\
&=&\sum_{1\leq i\leq N}(C^e_iE_i+C^d_iS_i+C_sS_i)\label{eq:cost-const}\\
&where&\mathbf{E}=[E_1,\cdots,E_N],~\mathbf{S}=[S_1,\cdots,S_N],\nonumber\\ 
& &\mathbf{1}=[1,\cdots,1]\label{eq:list}
\end{eqnarray}

In Eq.~(\ref{eq:cost-const}), these three dot products $\mathbf{C_e}\cdot\mathbf{E}$, $\mathbf{C_d}\cdot\mathbf{S}$, and $C_{s}\mathbf{1}\cdot\mathbf{S}$ represent the energy cost to apply DD, the data transfer price of distilled dataset, and the storage consumption cost, respectively. 
Eq.~(\ref{eq:list}) lists the DD energy ($\mathbf{E}$) and the data volume to transfer ($\mathbf{S}$) for $N$ edge devices. 
Here, $E_i$ denotes the DD energy consumed at the $i$th edge, while $S_i$ is equal to the size of the distilled dataset $\mathrm{D_i}$ generated at the $i$th edge node, both functions of $\mathrm{O_i}$, $\alpha_i$, and $\beta_i$. 



Next, we model the first term, i.e., the energy overhead of DD on edge devices empirically:
\begin{eqnarray} 
E_i&=&E_{dd}(\mathrm{O_i}, \alpha_i, \beta_i) \nonumber\\
&=& (d_{1,i} \alpha_i^2 + d_{2,i} \alpha_i + d_{3,i}) \cdot \beta_i + d_{4,i}
\label{eq:energy1}
\end{eqnarray}
As mentioned in Section~\ref{trade-off}, DD is typically implemented as an iterative solver. Consequently, the execution time and energy consumption of the main loop can be considered a linear function of $\beta_i$, the iteration count. 
Here, the intercept $d_{4,i}$ can be interpreted as the initialization overhead before the main loop. 
As for the parameter $\alpha_i$, we approximate its impact on per-loop runtime or energy using the quadratic function (see also Section~\ref{trade-off}), which is validated later in Section~\ref{subsub:validate energy prediction}. 

For the size function, we apply the simple analytical model described in Section~\ref{sec:dd-concept}. More specifically, the size of a distilled dataset at the $i$th edge ($\mathrm{D_i}$) is formulated as follows: 
\begin{eqnarray}
S_i=Size(\mathrm{D_i}=DD(\mathrm{O_i}, \alpha_i, \beta_i)) = KV\alpha_i\label{eq:size-dd}
\end{eqnarray}
Here, $K$ is the number of classes, while $V$ is the size of each data in the dataset. 
Note, $DD()$ represents the DD procedure at $i$th node, i.e., returning the distilled dataset $\mathrm{D_i}$ for a given input set of $\mathrm{O_i}$, $\alpha_i$, and $\beta_i$.


\subsubsection{\bf Statistical Accuracy Modeling}\label{subsub:Statistical Accuracy Modeling}
To constrain training quality, we model the expected test accuracy of the downstream deep learning task trained with the aggregated set of distilled datasets ($\mathbf{D}= \mathbf{DD}(\mathbf{O}, \boldsymbol{\alpha}, \boldsymbol{\beta})$). 
While several early studies have also attempted to model the test accuracy~\cite{accuracy-prediction1,accuracy-prediction2},
our work newly considers the variation of hyperparameter configurations uniquely set across $N$ devices in the modeling.
To this end, we treat the variation of hyperparameters $\boldsymbol{\alpha}, \boldsymbol{\beta}$ statistically and macroscopically for the following reasons: 

\begin{itemize}[noitemsep, labelindent=0pt, leftmargin=*]
\item The test accuracy is permutation invariant with respect to the elements in each hyperparameter vector $\boldsymbol{\alpha}$ / $\boldsymbol{\beta}$ particularly for IID data (see Section~\ref{sec:non-iid} for non-IID cases). 
\item For a permutation-invariant function $f(\mathbf{x})$, there exist functions $\rho$ and $\phi$ that meet $f(\mathbf{x}) = \rho(\sum_{x\in\mathbf{x}}\phi(x))$~\cite{zaheer2017deep}. 
\item Any sample moment can be expressed with the form of $\sum_{x\in\mathbf{x}}\phi(x)$, and the first, second, third, fourth, $\cdots$ order sample moments correspond to descriptive statistics, namely mean, variance, skewness, kurtosis, $\cdots$~\cite{moment}. 
\item After eliminating irrelevant statistics, a polynomial function of the remaining descriptive statistics should well represent the test accuracy in manner of Maclaurin expansion. 
\end{itemize}

In this work, we stop at the second order moment, and apply the following second order polynomial form by using  means ($\mu_{\alpha}, \mu_{\beta}$) and standard deviations ($\sigma_{\alpha}, \sigma_{\beta}$) of $\boldsymbol{\alpha}$ and $\boldsymbol{\beta}$: 
\begin{eqnarray}
Acc(\mathbf{D})
&\simeq& a_0 + a_1\mu_{\alpha}+a_2\mu_{\beta}+ a_3\mu_{\alpha}^2+a_4\mu_{\alpha}\mu_{\beta}\nonumber\\
& &+ a_5\mu_{\beta}^2+a_6\sigma_{\alpha}^2+a_7\sigma_{\beta}^2\label{eq:acc-model}\label{eq:accuracy-model}
\end{eqnarray}
This simple form of the test accuracy model is beneficial in limiting the calibration overhead and scaling the system because the number of unknowns is constant as the edge count $N$ scales. 
At the same time, limiting the orders of moment and polynomial contributes to avoiding overfitting, particularly compared to a model that fully uses all parameters in $\boldsymbol{\alpha}$ and $\boldsymbol{\beta}$ with a high polynomial order. 
Note that one can improve the accuracy of this function simply (1) by using additional descriptive statistics (e.g., skewness and kurtosis) or (2) by increasing the order of the polynomial approximation. 
Nevertheless, we use the above simple form based on our preliminary experiments, which is also validated later in Section~\ref{sec:order-select}. Consequently, it is lightweight but accurate enough for various edge counts $N$ as demonstrated later in Section~\ref{result}.


%

\begin{figure}[t]
\begin{center}
  \includegraphics[width=\linewidth]{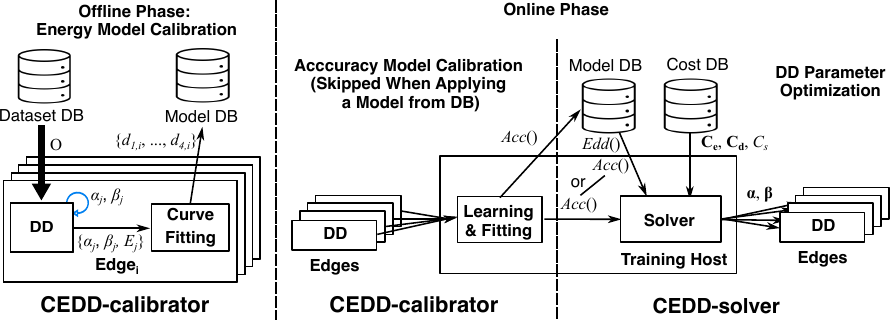}
  \caption{Overview of CEDD-optimizer Workflow}
  \label{fig:solution-arch}
\end{center}
\vspace{-10pt}
\end{figure}

\section{System Design}\label{solution}

\subsection{Solution Overview}\label{solution-overview}
Fig.~\ref{fig:solution-arch} illustrates the overall workflow of CEDD-optimizer to solve the optimization problem presented in Section~\ref{formulation} by using the model introduced in Section~\ref{modeling}. 
As shown, our solution consists of two modules: (1) the CEDD-calibrator and (2) the CEDD-solver. 
The CEDD-calibrator deals with online or offline model parameter identification, while the CEDD-solver optimizes the hyperparameter setup online. 

In the offline phase, the CEDD-calibrator runs the DD code on a benchmark dataset while varying the hyperparameters $\alpha$ and $\beta$, measures the energy consumption per iteration, and fits the polynomial curve described by Eq.~(\ref{eq:energy1}).
Note that this model calibration procedure is required \textbf{only once} per edge device. 
The calibration comprises two steps: measuring $d_{4,i}$ (initialization and finalization overhead) by setting $\beta=0$; and exploring $\alpha$ within $[\alpha_{min}, \alpha_{max}]$, while fixing $\beta$ at 1.


In the online phase, both modules are involved. The CEDD-solver solves the optimization problem by using the price parameters ($\mathbf{C_e}$, $\mathbf{C_d}$, $C_s$) stored in the cost database. 
In our implementation, we use \texttt{scipy.optimize} in \texttt{SciPy} with \texttt{SLSQP} option to solve the problem~\cite{scipy-optimizer}. 
Our solver requires the coefficients of the energy and test accuracy models as inputs. 
For the former, we use the coefficients identified in the offline calibration mentioned above, while  
for the latter, we offer two options to cope with the dependency of test accuracy on dataset features. 
The default option is reactive, agnostic to the dataset --- it learns the model coefficients online using a three-step tuning algorithm, described later in Section~\ref{subsection:three-step}. 
After this online model calibration is completed, the learned coefficients are stored in the model database for record or potential future reuse. 
Another option is proactive, using prior knowledge provided by the user --- it applies a user defined or pre-trained test accuracy model stored in the model database (e.g., reuse a model from a previous run). 
The approach is valid as long as the user is confident with the coefficient setup and reduces the overhead by skipping the online training. 


\begin{figure}[t]
\begin{center}
  \includegraphics[width=\linewidth]{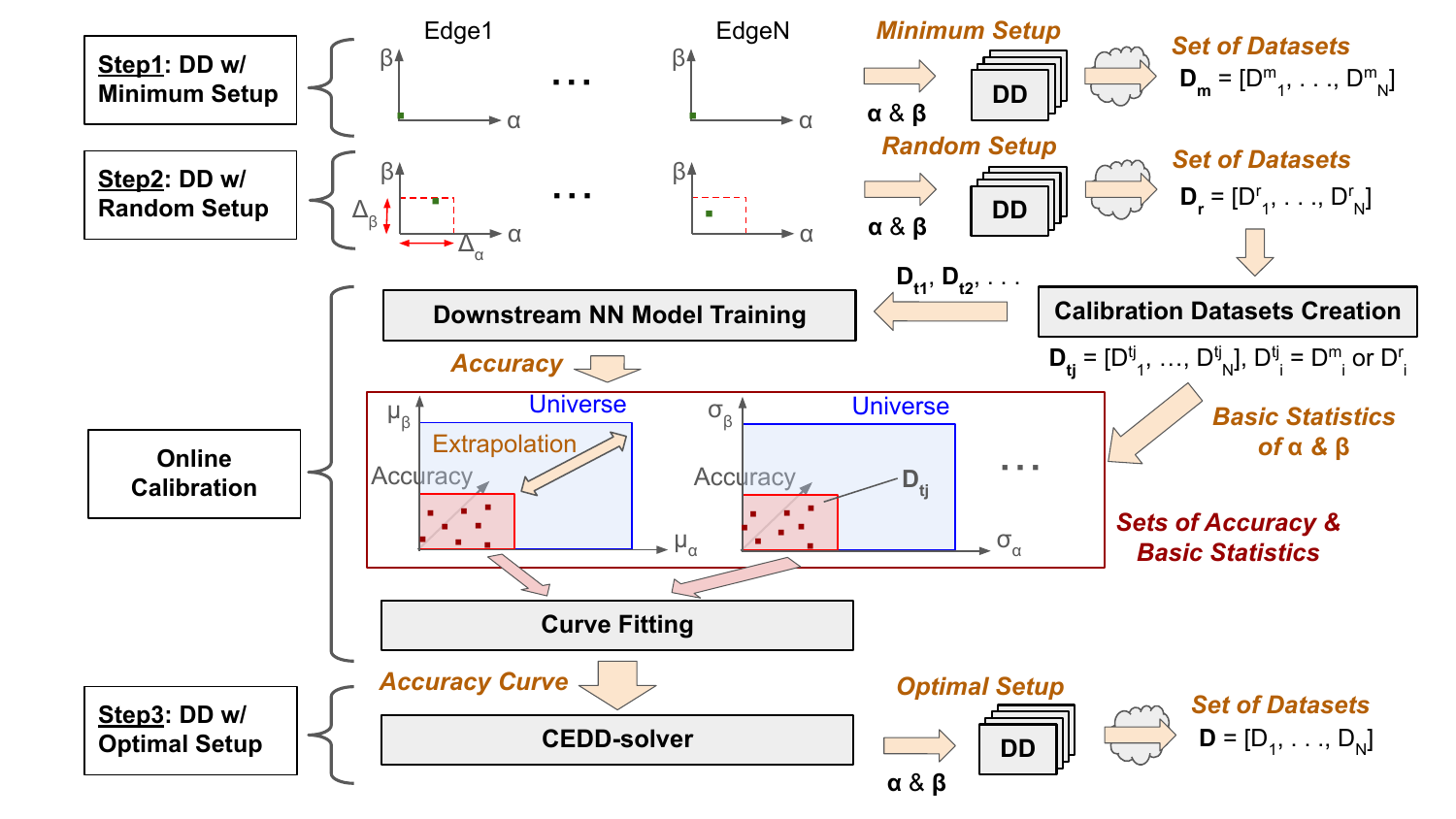}
  \caption{Online Three-Step Hyperparameter Tuning}
  \label{fig:3step-opt}
\end{center}
\vspace{-10pt}
\end{figure}

\subsection{Three-Step Hyperparameter Tuning Scheme}\label{subsection:three-step}
Fig.~\ref{fig:3step-opt} illustrates the online three-step tuning approach that coordinates the CEDD-calibrator and the CEDD-solver. 
Overall, the scheme aims to efficiently generate a training dataset that fits the test accuracy curve. 

In the first step, DD is applied across all edges while setting $\alpha$ and $\beta$ to the minimum, i.e., $\alpha_{min}=1$ and $\beta_{min}=1$, respectively. 
Then, a set of distilled datasets $\mathbf{D_m}=\left[\mathrm{D^m_1}, \cdots, \mathrm{D^m_N}\right]$ is generated where $\mathrm{D^m_i}$ represents the dataset distilled at the $i$th edge node. 
As the energy consumption of DD is modeled as $O(\alpha^2\beta)$, as denoted in Eq.~(\ref{eq:energy1}), the overhead of this step is almost negligible.

The second step then creates $N$ sets of $(\alpha, \beta)$ where $\alpha$ and $\beta$ are randomly chosen from $\left[\alpha_{min}, \alpha_{min}+\Delta_\alpha\right]$ and $\left[\beta_{min}, \beta_{min}+\Delta_\beta\right]$, respectively. 
Note that the total cost, including the DD energy consumption and the subsequent data transfer, is controlled by $\Delta_{\alpha}$ and $\Delta_{\beta}$. 
Once the set of $(\alpha, \beta)$ is generated, its mapping to $N$ edge devices is optimized to minimize the total cost, encompassing both energy consumption and data transfer across $N$ edge devices, by solving the minimum-weight perfect bipartite matching problem (MWPBM). 
Then, an assignment solver namely \texttt{linear\_sum\_assignment()}~\cite{scipy-assignment} in \texttt{SciPy} is used, which solves the problem using the Jonker-Volgenant algorithm~\cite{crouse2016implementing}. 
Consequently, a set of distilled datasets $\mathbf{D_r}=\left[\mathrm{D^r_1}, \cdots, \mathrm{D^r_N}\right]$ is collected ($\mathrm{D^r_i}$:~the distilled dataset at the $i$th edge node). 

Once the datasets $\mathbf{D_m}$ and $\mathbf{D_r}$ are collected at the host node, multiple training datasets are created $\mathbf{D_{t_1}}, \mathbf{D_{t_2}}, \cdots$ where $\mathbf{D_{t_j}}=\left[\mathrm{D^{t_j}_1}, \cdots, \mathrm{D^{t_j}_N}\right]$ ($\mathrm{D^{t_j}_i} = \mathrm{D^{m}_i}~or~\mathrm{D^{r}_i}$, $i$: edge ID). 
Here, the $j$th training dataset $\mathbf{D_{t_j}}$ is composed by choosing the distilled dataset generated either in the first or second step ($\mathrm{D^{m}_i}$ or $\mathrm{D^{r}_i}$) for each edge $i$, which is randomly selected. 
The set of $\alpha$ and $\beta$ setups across $N$ edges is associated with $\mathbf{D_{t_j}}$, and the basic statistics (e.g., $\mu_{\alpha}$) are calculated. 

Then, the training host performs the downstream neural network training and outputs the test accuracy for all training datasets $\mathbf{D_{t_1}}$, $\mathbf{D_{t_2}}$, $\cdots$. 
As these datasets are distilled and, hence, very compact, the overhead of this training is very limited as demonstrated later in Section~\ref{result} (4th graph in Fig~\ref{fig:experiment-compilation}). 
We then combine the test accuracy output and the basic statistics for each training dataset $\mathbf{D_{t_j}}$. 
We fit the test accuracy curve modeled in Section~\ref{subsub:Statistical Accuracy Modeling} to the combinations of test accuracy and basic statistics across all training datasets
for interpolation and extrapolation. 
To guide this fitting procedure, predefined boundary information is used.
Note that the accuracy of this test accuracy curve is controllable by $\Delta_{\alpha}$ and $\Delta_{\beta}$. 

Finally, CEDD-solver optimizes the setups of $\boldsymbol{\alpha}$ and $\boldsymbol{\beta}$ by using the test accuracy model, the energy model, and the price parameters, then DD is performed across edge devices, with the optimal hyperparameter setting. 
Our scheme takes three or less steps to find the optimal setting, while limiting the overhead spent on the first two steps. 
More specifically, ours can stop at the end of the first or second step if the test accuracy meets the target $A_{trg}$ for $\mathbf{D_m}$ or $\mathbf{D_r}$. 


\section{Evaluation}\label{evaluation}

\subsection{Evaluation Setup}\label{methodology}

This section describes the details of our evaluation setup. The default parameter setup is listed in Table~\ref{table:default-params}, but we adjust them for each experiment based on the experiment purpose, as documented in the relevant sections. 

\begin{table}[b]
\caption{Default Parameter Setup for DD}
\label{table:default-params}
{
\footnotesize
\centering
\begin{tabular}{|M{0.95\linewidth}|}
    \hline
    Parameter Setup \\
    \hline\hline
    $\alpha_{min}$=$\beta_{min}$=$1$, $\alpha_{max}$=$70$, $\beta_{max}$=$40$, $\Delta{\alpha}$=$25$, $\Delta_{\beta}$=$15$, $N$=$10$, $R_{trg}$=$ M/(NK  \alpha_{max})$, dataset = MNIST~\cite{mnist}, host = MacBook, edges = TQMx80UC x2, MacBook x3, Jeston Orin x2, Jetson Xavier x3\\
    \hline
\end{tabular}
}
\end{table}



\subsubsection{\bf Hardware Platforms}\label{subsub:hw}
We focus on the following devices: (1) NVIDIA Jetson Xavier NX; (2) NVIDIA Jetson Orin Nano; (3) MacBook Pro; and (4) TQ module TQMx80UC. 
The first two are embedded computing platforms designed for edge and AI computing. The Xavier NX comprises a 6-core ARM Carmel CPU and a 384-core Volta GPU, 
while the Orin Nano is composed of 6-core ARM Cortex-A78AE CPU and an Ampere GPU.  
Both are representatives of edge‑AI computing systems, powered by Tensor Cores that enable energy-efficient training and inferences for deep learning tasks. 
The MacBook Pro Quad‑Core is a powerful mobile computing platform, consisting of an Intel Core-i5 processor and a discrete GPU. 
The TQMx80UC COM module~\cite{tq-system} is an x86-based edge computing platform, comprising a long‑lifecycle Intel core processor (8th generation). 
These systems cover a wide spectrum of edge and mobile computing platforms, are suitable for our target scenario, and ensure the robustness of our experiments with respect to hardware characteristics.

\subsubsection{\bf Baseline Datasets and Networks}\label{subsub:Datasets and Parameterized DD}
We use publicly available standard datasets to illustrate the problem in a controlled and reproducible manner:  MNIST~\cite{mnist}, Fashion-MNIST~\cite{F-mnist}, CIFAR-10~\cite{cifar10}, SVHN~\cite{SVHN}, ImageNette~\cite{ImageNette} (subset images of ImageNet \cite{ImageNet}). 
The applicability to even larger datasets is discussed in Section~\ref{scale}. 
While we use these image datasets following the literature, very recent studies report that DD is also applicable to time series data~\cite{hong2026harmonic}. 
We evaluate our methodology under two scaling regimes, in terms of the number of edge devices $N$: \textit{Strong Scaling}, i.e., the total dataset size remains fixed while the number of edge devices increases; and \textit{Weak Scaling}, i.e., the dataset size per edge node remains constant when the number of edge devices scales.

We focus on three data-distribution settings:(1) Independent and Identically Distributed (IID) data; (2) quantity-based near-IID data generated using Dirichlet distribution with concentration parameter of $\alpha=1.0$; (2) quantity-based non-IID data generated using Dirichlet distribution with concentration parameter of $\alpha=0.5$.
In all settings, we ensure that each edge device contains at least 15 samples from each class.
For our test accuracy model, the IID setup is more suitable by its design. 
A potential extension is discussed later in Section~\ref{sec:non-iid}. 

For our downstream deep learning tasks, we use a neural network architecture designed by Gidaris and Komodakis~\cite{network}. The architecture encompasses (1) three convolutional blocks with 3x3 filters, (2) instance normalization~\cite{Instance-norm}, (3) RELU activation, and (4) 2×2 average pooling with a stride of 2. 

\begin{figure*}[t]
        \centering
        \includegraphics[width=\linewidth]{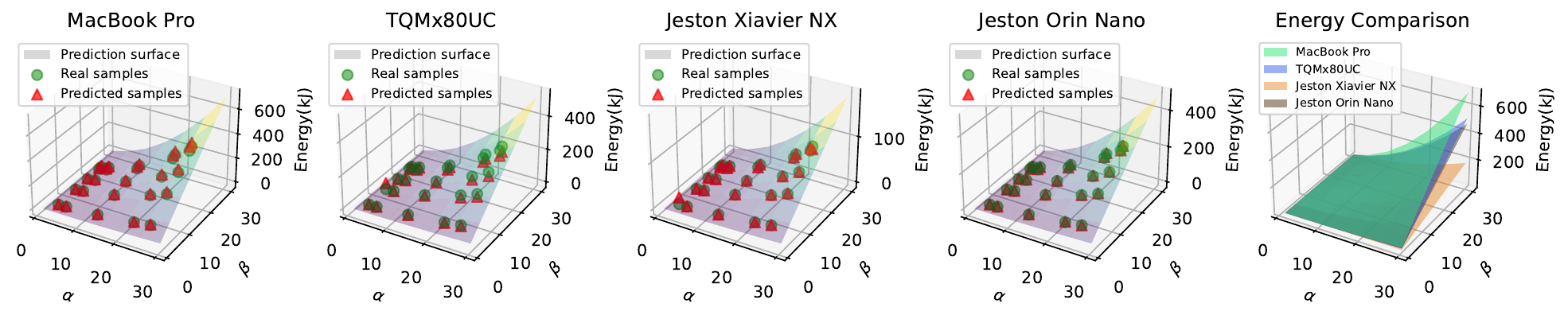}
        \caption{Time and Power Measurement and Their Approximation on Various Devices}
        \label{fig:energy-curves-ablation}
        \vspace{-2pt}
\end{figure*}

\subsubsection{\bf DD Implementation}\label{subsub:Parameterized DD}
Among the DD methods introduced in Section~\ref{sub:dd-literature}, we select the vanilla Dataset Condensation (DC)~\cite{dd-dc} as our DD baseline. 
This is because it exhibits fundamental structures commonly found in most other mainstream studies and implementations~\cite{dd-dsa,dd-dm,dd-cafe,dd-mae,dd-mtt,dd-att}. 
As we apply a parametrization to the commonly available basic structure, our approach is broadly applicable to other DD frameworks that may integrate sophisticated approaches/concepts, such as improved network architectures~\cite{dd-kip,dd-gan}, label learning~\cite{dd-sl,dd-sl1,dd-tasla}, long-range trajectories~\cite{dd-mtt,dd-att}, and others~\cite{dd-cafe,dd-prunning,dd-dm,dd-attacks1,dd-other7}. 

\subsubsection{\bf Energy Model Calibration}\label{subsub:energy model detail}
To identify the coefficients of our energy model in Eq.~(\ref{eq:energy1}), we measure the average runtime and power while scaling DPC ($\alpha$) selected from $\{1, 10, 20, 30, 40, 50, 60, 70\}$. 
We measure power consumption using the following tools: \texttt{tegrastats} for Jetson Xavier NX and Jetson Orin Nano; \texttt{powermetrics} for MacBook Pro; and \texttt{powerstat} for the TQMx80UC. 
Note that the clock frequency is set to the default for all, and coordination with clock scaling in our approach 
is orthogonal to our optimization, as discussed later in Section~\ref{discussion}.
For the energy calibration, we divide the DD execution into two phases as mentioned in Section~\ref{solution-overview}: (a) the initialization (i.e., encompassing module loading and data initialization), and (b) the main loop of the DD iterative solver.

\subsubsection{\bf Price Parameter Setup} \label{subsub:cost setups}
The energy and size models are then used to quantify the total cost formulated in Eq.~(\ref{eq:cost-const}). 
To this end, the price parameters $\mathbf{C_e}$, $\mathbf{C_d}$, and $C_s$ need to be identified beforehand and depend heavily on the geographical locations of edge servers, as previously mentioned. 
For the energy price parameters $\mathbf{C_e}$, we use the same data presented in Fig.~\ref{fig:electricity-cost}, which is based on the World Population Review~\cite{electricity}\footnote{Although the electricity price depends also on the contract and varies within a country, the parameter setups still covers the real-world prices. }. 
As for the network price parameters $\mathbf{C_d}$, we extract the operating cost per region (SiteLink Rate) from~\cite{aws-price}, which is the same as Fig.~\ref{fig:sitelink_plot}. 
Regarding the storage price parameter, we model it after Amazon's S3 pricing~\cite{storage-price} and set it to \$ 0.023 per GB, regardless of the edge locations. 

As for the geographical location setting, we first choose 74 out of the 144 countries listed in the World Population Review~\cite{electricity} based on the applicability of the list shown in Fig.~\ref{fig:sitelink_plot}. 
We then set the geographical location for each of the $N$ edge devices by randomly choosing one of the 74 countries.

\subsubsection{\bf Baseline Methods}
We compare against a DD baseline with $(\alpha,\beta)=(\alpha_{max}, \beta_{max})$ and conventional random sampling~\cite{dd-dc,dd-dm,dd-mtt}. 
We further compare ours with commonly-used surrogate-free iterative search algorithms (i.e., random search, heuristic search, exhaustive search~\cite{hearisticsearch}). 
The advantage of DD over generic lossy/lossless compression methods is shown in Fig.~\ref{fig:DD-prove3}. Note, the compression methods either do not satisfy the target compression ratio $R_{trg}$ or significantly violate the target accuracy $A_{trg}$. Consequently, we do not include them in our comparisons.



\subsection{Experimental Result}\label{result}



\subsubsection{\bf Validation of the Energy Model}\label{subsub:validate energy prediction}
We perform our energy model experiments on the following platforms: MacBook Pro, TQMx80UC, Jeston Xiavier NX, and Jeston Orin Nano using the setup detailed in Section~\ref{subsub:energy model detail}. 
Fig.~\ref{fig:energy-curves-ablation} visualizes the ground truth, fitted surface, and predicted values for these four platforms as well as the comparison of energy surfaces among the platforms. 
Overall, the energy model matches the ground truth accurately, as the predicted values are very close to the ground truth energy. 
The prediction errors are generally very small for all devices --- lower than \textbf{3\%}. 
As shown in the rightmost figure, the shapes of energy surfaces differ significantly across devices, underscoring the importance of hardware-aware hyperparameter tuning when the system is heterogeneous in terms of edge hardware. 
Our tuning approach covers this aspect by simply using device-specific coefficients ($d_{1,i}$, $\cdots$, $d_{4,i}$) for each edge node.

\begin{figure}[t]
\begin{center}
  \includegraphics[width=\linewidth]{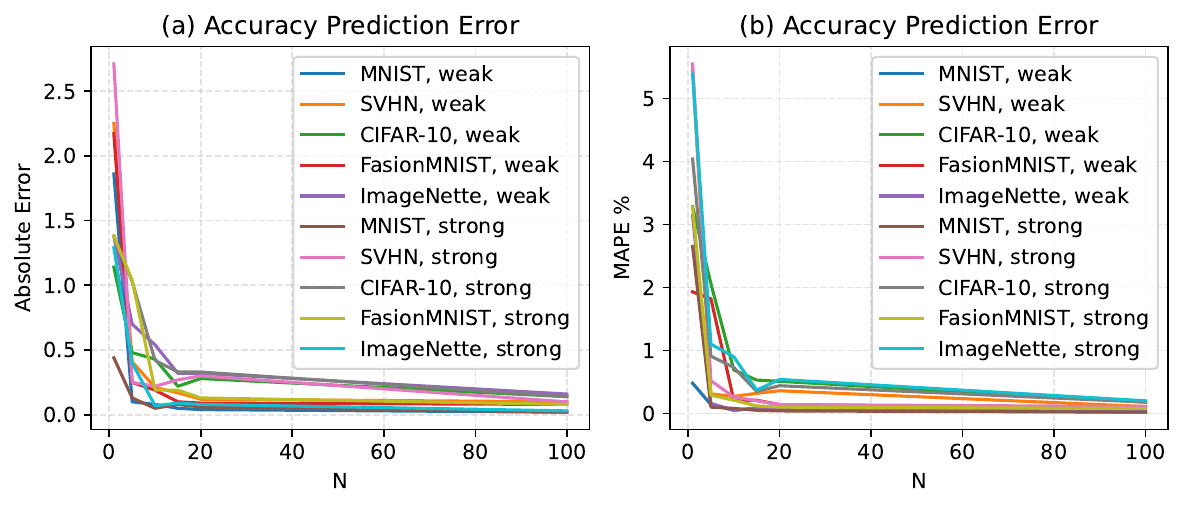}
        \caption{Error v.s. Scale in Our Test Accuracy Model $Acc()$}
        \label{fig:weak strong}
\end{center}
    \centering
    \includegraphics[width=\linewidth]{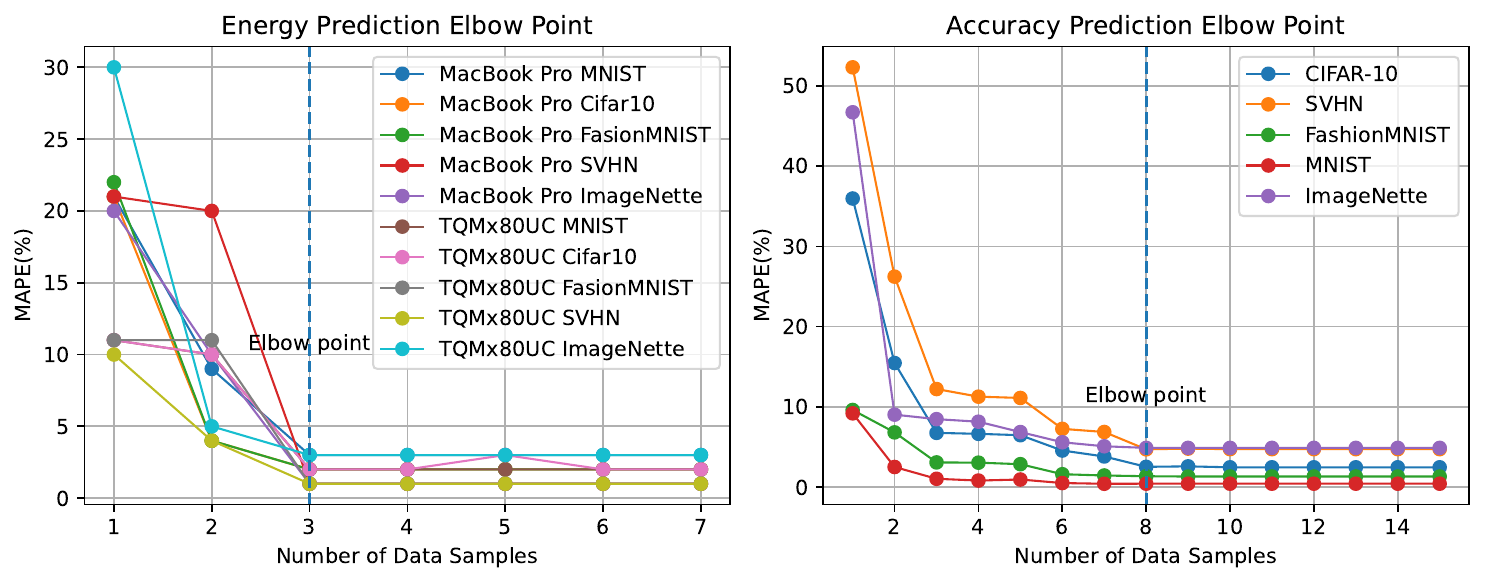}
    \caption{Model Error as a Function of Fitting Sample Count}
    \label{fig:error-vs-sample-count}
\end{figure}

\begin{figure*}[t]
    \centering
    \includegraphics[width=\linewidth, trim=85 12 85 25, clip]{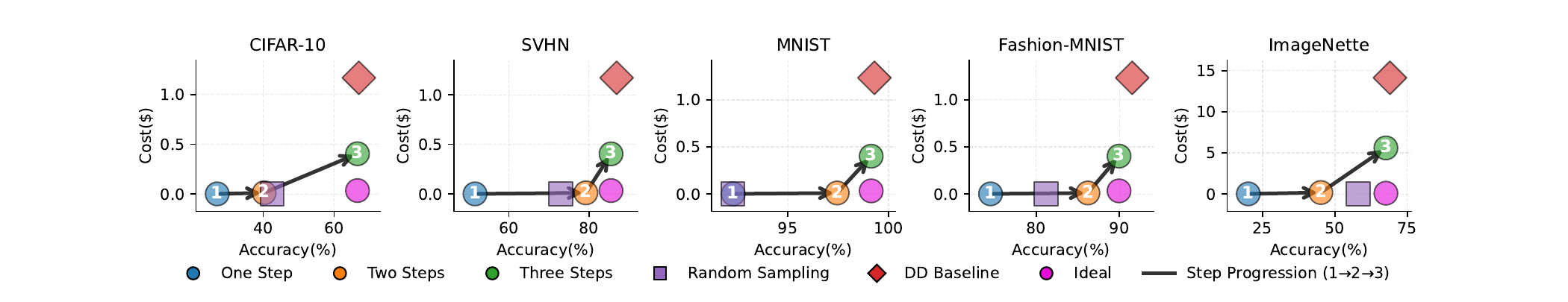}
    \caption{Cost and Test Accuracy Comparison across Various Standard Datasets}
    \label{fig:cost-accuracy-comparison}
\end{figure*}

\subsubsection{\bf Validation of Accuracy Model}\label{subsub:validate accuracy prediction}
We estimate the test-accuracy model coefficients as follows.
First, we randomly generate 18 sets of $\boldsymbol{\alpha} = [\alpha_1, \alpha_2, \cdots, \alpha_N]$ and $\boldsymbol{\beta}= [\beta_1, \beta_2, \cdots, \beta_N]$.
For each hyperparameter setup, we run DD on every edge server, aggregate the distilled data centrally, and train/evaluate a randomly initialized network on the aggregated set. This yields 18 test accuracies $Acc$, each averaged over five variants, together with the corresponding descriptive statistics ($\mu_{\alpha}$, $\cdots$).
Note, we found that 18 samples are sufficient for stable performance based on a pilot study with up to 100 samples. We split the data into 8 training and 10 test points for fitting Eq.~(\ref{eq:acc-model}) using stochastic gradient descent (SGD)~\cite{sgd1,sgd2}, yielding coefficients $a_0,a_1,\ldots,a_7$. 


We present the errors of our test accuracy model ($Acc()$) using various datasets while scaling the number of edge nodes $N$ by using the setup detailed in Section~\ref{subsub:Statistical Accuracy Modeling}. 
Fig.~\ref{fig:weak strong} shows the result for both weak and strong scaling (defined in Section~\ref{subsub:Datasets and Parameterized DD}): (a) absolute error and (b) Mean Absolute Percentage Error (MAPE).
For MNIST~\cite{mnist}, FashionMNIST~\cite{F-mnist}, SVHN~\cite{SVHN} and CIFAR‑10~\cite{cifar10}, the absolute error and MAPE are smaller than \textbf{0.5} and \textbf{1\%}, respectively, when the number of edges $N$ is greater than 10. 
\textit{Generally, the error becomes smaller as we scale $N$, and saturates at a certain point. }
We assume this behavior is caused by its statistical nature (i.e., the central limit theorem) as we utilize several descriptive statistics such as the means in the test accuracy model ($Acc()$). 

\begin{figure*}[t]
    \centering
    \includegraphics[width=\linewidth]{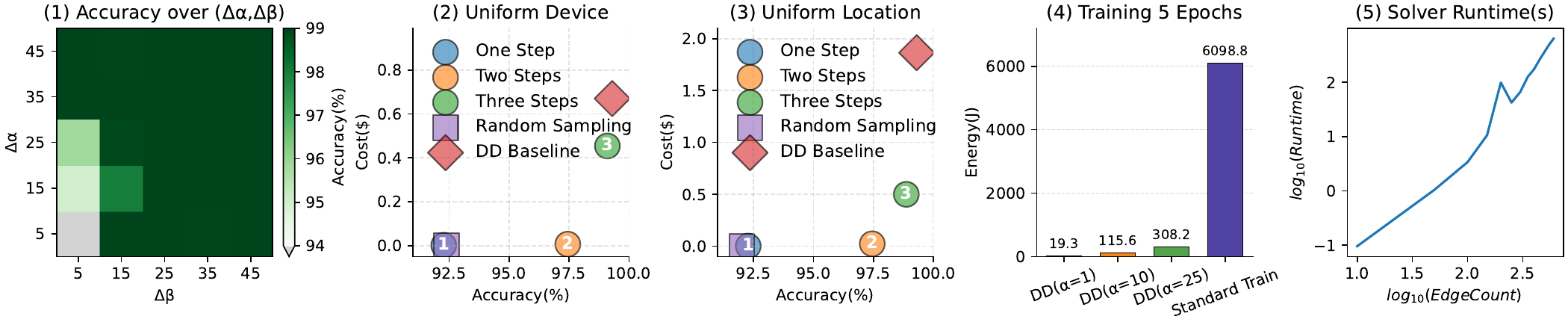}
    \caption{A Compilation of Experimental Results: (1) Impact of $\Delta_{\alpha}$ and $\Delta_{\beta}$ on Accuracy of Test Accuracy Model; (2)/(3) Test Accuracy v.s. Cost for Uniform Device/Location Setup; (4) Comparison of Energy Overhead for Downstream Training Task; and (5) Solver Runtime as a Function of Edge Count $N$}
    \label{fig:experiment-compilation}
\end{figure*}

\subsubsection{\bf Sample Count Selection in CEDD-calibrator}
Fig.~\ref{fig:error-vs-sample-count} presents the error as a function of the number of sampling points to fit for our energy and test accuracy modeling. 
As an error metric, we again use MAPE. 
The X-axis indicates the number of samples, while the Y-axis represents MAPE~[\%]. 
Our accuracy and energy models achieve MAPEs below 6\% and 3\%, respectively, across various datasets. 
Note, an MAPE below 10\% is generally considered a good fit. 
Based on this experiment, we judge that 8 and 3 fitting points are sufficient for our accuracy or energy models, respectively. 
In our CEDD-Calibrator implementation, we sample more to set margins and test the curves after the fitting. 
Consequently, we sample 18 or 5 points for the accuracy or energy models, respectively.


\subsubsection{\bf Cost and Test Accuracy Benefits by CEDD-optimizer}\label{sub:ablation on parameters}
Fig.~\ref{fig:cost-accuracy-comparison} showcases the effectiveness of our CEDD-optimizer across 5 standard datasets. 
The horizontal axis indicates the accuracy, whereas the vertical axis indicates the total cost encompassing DD energy consumption, data transfer, cloud storage, and host energy consumption. 
Intuitively, the lower right region is better on the 2D plane. 
In this experiment, we compare the following methods in terms of cost and accuracy: (a) \texttt{One Step}, finishing the tuning after the first step in CEDD-optimizer ($\alpha=\beta=1$); (b) \texttt{Two Step}, finishing the tuning after the second step; (c) \texttt{Three Step}, performing the whole procedures in our three step tuning; (d) \texttt{Random Sampling}, randomly choosing a subset of the original dataset in each edge device such that the total data transfer size becomes the same as that of \texttt{Three Step} (the same size for all edges); (e) \texttt{DD Baseline}, applying DD at $\alpha=\alpha_{max}$ and $\beta=\beta_{max}$; and (f) \texttt{Ideal}, CEDD-optimizer when using the accuracy model from the database that is trained with $\Delta_{\alpha}=\alpha_{max}-\alpha_{min}$ and $\Delta_{\beta}=\beta_{max}-\beta_{min}$ (ideal scenario). 
Overall, our three-step tuning scheme significantly reduces the cost compared with the \texttt{DD Baseline}, while keeping almost the same accuracy.
As implied by the cost difference between \texttt{Three Step} and \texttt{Ideal}, one can achieve an even further cost reduction by reusing the test accuracy model, if applicable. 
Compared with \texttt{Random Sampling}, \texttt{Ideal} improves accuracy significantly while keeping the cost almost the same, implying that tuning is as simple as \texttt{Random Sampling}. 
CEDD-ideal reduces cost by $5.25\times$--$20.8\times$ and the three-step algorithm by $1.7\times$--$2.67\times$ versus the DD baseline, both with only a $1\%$--$1.5\%$ accuracy drop.

\subsubsection{\bf Impact of $\Delta_{\alpha}$ and $\Delta_{\beta}$ on Accuracy Model Quality in Our Three Step Tuning}
The effectiveness of our three-step tuning scheme depends on the setup of $\Delta_{\alpha}$ and $\Delta_{\beta}$. 
The lower $\Delta_{\alpha}$ and $\Delta_{\beta}$ are set, the less cost the calibration requires, but also the less accurate the test accuracy model becomes, 
as these parameters control the ranges of $\alpha$ and $\beta$ selected in the second step (see Section~\ref{subsection:three-step}. 
As $\Delta_{\alpha}$ and $\Delta_{\beta}$ are set smaller, the covered regions of basic statistics (e.g., $\mu_{\alpha}$) also become smaller, 
which degrades the test accuracy function's accuracy as a consequence. 
The leftmost heatmap in Fig.~\ref{fig:experiment-compilation} visualizes the impact of $\Delta_{\alpha}$ and $\Delta_{\beta}$ on the accuracy of the test accuracy model. 
The gray region ($\Delta_{\alpha}=\Delta_{\beta}=5$) indicates a failure case, in which the solver does not find a solution. 
Based on the observation, we set $(\Delta_{\alpha},\Delta_{\beta})=(25,15)$, which well balances the trade-off throughout our experiments.

\subsubsection{\bf Sensitivity to Device/Location Heterogeneity}
The plots numbered (2) and (3) in Fig.~\ref{fig:experiment-compilation} demonstrate our two ablation studies conducted for \texttt{MNIST}, i.e., removing the heterogeneity in terms of edge hardware and location from the default. 
More specifically, we use the same edge device, namely \texttt{Jetson Xavier}, for the former, while the device location is set uniformly for the latter. 
The other evaluation parameters are unchanged from the default. 
Overall, our three-step tuning scheme is still effective compared with the \texttt{DD Baseline} and \texttt{Random Sampling} for these less heterogeneous environments. 


\subsubsection{\bf Cost Reduction by DD for Downstream Learning}
Our three-step tuning scheme requires multiple downstream training runs to generate samples for fitting the test-accuracy curve. 
However, this overhead is limited because downstream training is performed on distilled datasets.
Fig.~\ref{fig:experiment-compilation} (4) shows this effect using datasets distilled under different hyperparameter setups
The vertical axis reports the training energy consumption over five epochs, while the horizontal axis lists the downstream training with different datasets. 
Specifically, \texttt{DD($\alpha$=X)} shows dataset distilled with $\alpha=X$, whereas \texttt{Standard Train} utilizes the original dataset without DD with batch size equals 512.
Here, experiments are on \texttt{MNIST}. 
Since this significant training cost reduction in the downstream training task, 
Our three-step tuning scheme is effective as shown in Fig.~\ref{fig:cost-accuracy-comparison}. 


\begin{figure*}[t]
    \centering
    \includegraphics[width=\linewidth]{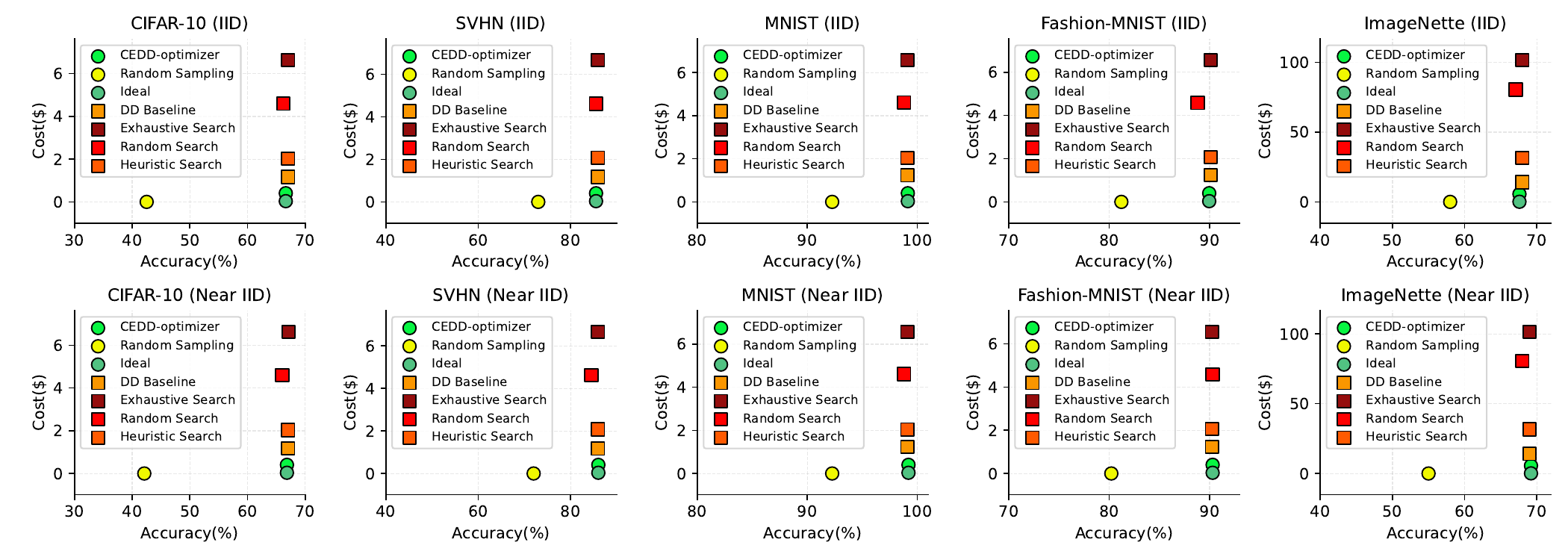}
    \caption{Accuracy-Cost Comparisons with Baselines and Iterative Search Methods under IID and Near-IID Scenarios}
    \label{fig:extend1}
\end{figure*}

\subsubsection{\bf Scalability Assessment}\label{scale}
In this section, we assess the scalability of our CEDD-optimizer in terms of cost by analyzing the scaling properties. 
Note that the scaling rule of the test accuracy error is presented and discussed in Section~\ref{subsub:validate accuracy prediction}. 
Here, we consider two scaling policies: weak and strong scaling. 
For the former, we assume that the total data count $M$ across all edge devices is scaled proportionally to the edge count $N$, while the number of classes $K$, the size per data $V$, and the distributions of $\alpha$ and $\beta$ are kept constant. 
As for the latter, we assume that the total data count $M$ remains constant as the edge device count $N$ scales.  
The other scaling conditions for $K$, $V$, $\alpha$, and $\beta$ are the same as those of weak scaling. 

\begin{table}[b]
\caption{Cost Scaling Rules in CEDD-optimizer}
\label{table:scalability}
{
\footnotesize
\centering
\begin{tabular}{|M{0.2\linewidth}||M{0.1\linewidth}|M{0.1\linewidth}|M{0.085\linewidth}|M{0.1\linewidth}|M{0.095\linewidth}|}
    \hline
    Scaling Policy & Total~DD Energy & Traffic~\& Storage & Fitting & Learning & \textbf{Solver} \\
    \hline\hline
    Weak w/ DD & $O(N)$ & $O(N)$ & $O(1)$ & $O(N)$ & $\boldsymbol{O(N^3)}$ \\
    \hline
    Weak w/o DD & – & $O(N)$ & – & $O(N)$ & – \\
    \hline
    Strong w/ DD & $O(N)$ & $O(N)$ & $O(1)$ & $O(N)$ & $\boldsymbol{O(N^3)}$ \\
    \hline
    Strong w/o DD & – & $O(1)$ & – & $O(1)$ & – \\
    \hline
\end{tabular}
}
\end{table}

Table~\ref{table:scalability} lists scaling factors for various costs in both scaling policies with or without DD. 
The scaling factors of "with DD" are derived as follows: 
First, as mentioned in Section~\ref{energy-formulation}, the energy overhead of DD per edge is $O(\alpha^2\beta)$, which is not a function of $N$, thus the total energy cost for DD is $O(N)$ for both scaling policies. 
Moreover, as the volume of the distilled dataset per node is denoted as $KV\alpha$ (see Section~\ref{energy-formulation}), the data transfer and storage costs are also $O(N)$ for both scaling policies. 
As the curve to fit in our approach is scale free (see Section~\ref{subsub:Statistical Accuracy Modeling}), the fitting overhead is $O(1)$ for both policies. 
As for the cost of the downstream learning, 
Assuming the number of training epochs scales proportionally with the total dataset volume while the batch size remains constant, the learning overhead scales as $O(N)$ under both scaling policies. 
However, the overhead of the solver for tuning hyperparameters is $O(N^3)$, and thus this can cause bottlenecks at large scale. 
In Fig.~\ref{fig:experiment-compilation}~(5) represents the solver's run time as a function of edge device count $N$. 
In our environment, the solver takes 3.4s at $N=100$, but it can become a bottleneck at larger scale. 
One prominent solution is simply 
to replace the algorithm (SLSQP) with a more efficient heuristic alternative.

Under weak scaling, the total cost (excluding the solver) scales as $O(N)$ both with and without DD. 
Consequently, the cost-reduction ratio relative to the ``without DD'' baseline remains $O(1)$, i.e., \textit{DD provides scale-invariant benefits under weak scaling}. 
Under strong scaling, the cost-reduction ratio decreases with $N$ by a factor of $O(1/N)$. 
In practice, we expect weak scaling to be more common, since deploying additional edges is typically driven by the need to collect more data.


\begin{figure*}[t]
    \centering
    \includegraphics[width=\linewidth]{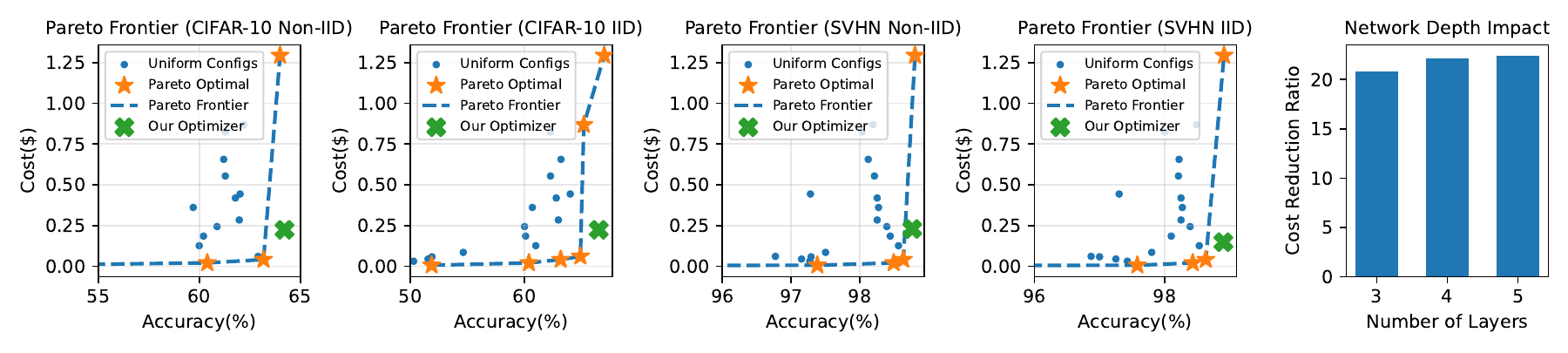}
    \caption{Our Optimizer with a Non-Uniform Hyperparameter Setup v.s. Uniform Hyperparameter Setups under IID and Non-IID Scenarios (Four Leftmost Graphs) and Impact of Network Depth on Cost Saving Factor (Rightmost)}
    \label{fig:extend2}
\end{figure*}

\begin{figure*}[t]
    \centering
    \includegraphics[width=0.7\linewidth]{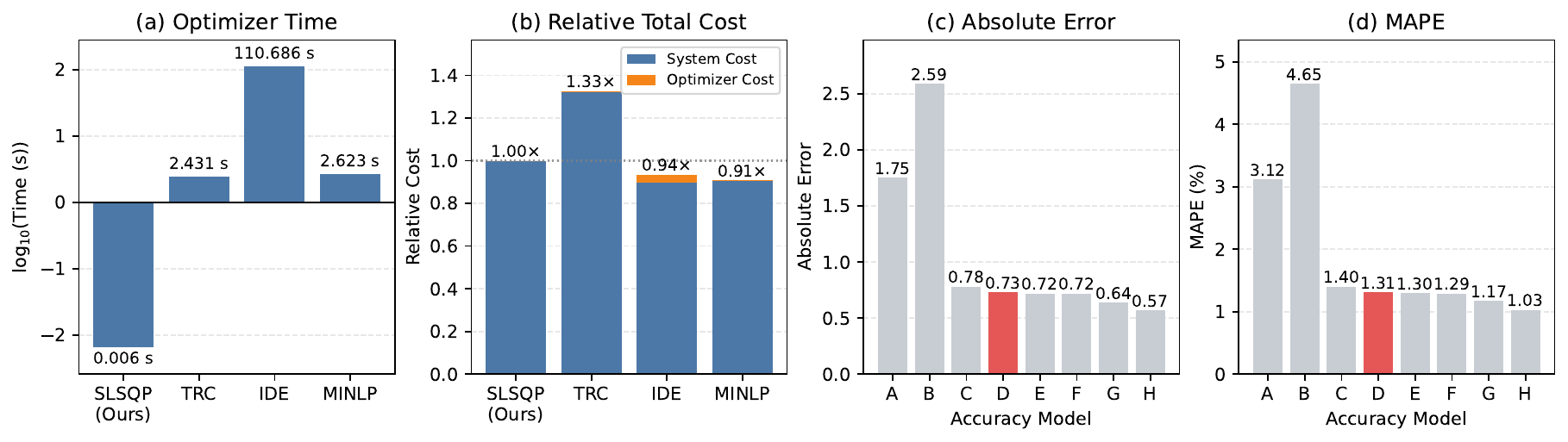}
    \caption{Time (a) and Cost (b) Comparison with Alternative Solvers and Test Accuracy Model Comparison in Absolute Error (c) and MAPE (d) }
    \label{fig:extend3}
\end{figure*}

\subsubsection{\bf Comparison to Iterative Search Methods}
It is hard for \texttt{Random sampling} to reach a near \texttt{Ideal} performance, while merely using \texttt{DD baseline} is also inefficient. 
Our three-step approach efficiently searches an optimal configuration, while intelligently limiting the search overhead. 
Fig.~\ref{fig:extend1} showcases its efficiency by comparing it with three commonly-used iterative search algorithms and the baselines.  
\texttt{Exhaustive search} is implemented as a grid search over $(\alpha\in{1,15,35,55,70})$ and $(\beta\in{1,10,15,20,30})$, while \texttt{Random search} iteratively evaluates 20 randomly generated ($\boldsymbol{\alpha}$, $\boldsymbol{\beta}$) configurations. 
\texttt{Heuristic search} starts from the smallest parameter values and increases $\alpha$ in steps of 15 and $\beta$ in steps of 10 until the accuracy constraint is satisfied.
These three methods iteratively search an optimal hypearparameter configuration in a step-by-step manner. 
Each iteration requires both the DD and the downstream training task, and the overhead depends on the choice of the hyperparameter vectors $\boldsymbol{\alpha}$ and $\boldsymbol{\beta}$. 

\texttt{Exhaustive search} and \texttt{Heuristic search} generally satisfy the constraint under both IID and near-IID data, whereas \texttt{Random search} can fail to find a feasible setup within the 20 attempts. 
Nevertheless, these three approaches are even more expensive than \texttt{DD baseline} that performs only one DD run with the maximum parameter configuration ($\alpha_i$, $\beta_i$)=($\alpha_{max}$, $\beta_{max}$). 
\texttt{CEDD-optimizer} intelligently avoids this situation by (1) restricting the search space with $\Delta_{\alpha}$ and $\Delta_{\beta}$ to control the overhead per search, (2) limiting the number of search iterations while fitting a simple curve, and (3) pinpointing an optimal with a solver after the fitting. 



\subsubsection{\bf Benefit of Non-Uniform Hyperparameter Setup}
Our CEDD-optimizer aims to configure the DD hyperparameters $\boldsymbol{\alpha}$ and $\boldsymbol{\beta}$ uniquely across edge devices depending on the heterogeneous price parameters. 
We then assess the benefit of this non-uniform tuning compared with uniform setups, in terms of the cost and the test accuracy. 
To do so, we test all the possible uniform $\boldsymbol{\alpha}$ and $\boldsymbol{\beta}$ settings, draw the Pareto frontier of the uniform setups on the accuracy-cost plane, and compare our non-uniform approach with it. 
Note that the same identical set of $\alpha$ and $\beta$ is applied to all edge nodes for the uniform setups, while $\alpha$ and $\beta$ can be different values. 

The four leftmost plots in Fig.~\ref{fig:extend2} compare our approach with the uniform setups and their Pareto frontier for \texttt{CIFAR-10} and \texttt{SVHN} under IID and non-IID scenarios. 
Across all four dataset distribution scenarios, our approach is better than the Pareto frontier. 
In practice, the benefit would be even more considerable as it would not always be possible to select a Pareto optimal setup, and suboptimal uniform setup points are mostly far from our optimizer on this plane.


\subsubsection{\bf Impact of Network Depth on Cost Saving}
The rightmost plot in Fig.~\ref{fig:extend2} evaluates how the network depth or size setup for DD affects the benefit of our method in the previously mentioned \texttt{Ideal} case. 
We evaluate this impact under the assumption that the test accuracy function remains unchanged for different network depths, in order to eliminate cost change factors other than the network depth, most prominently the hyperparameter setups $\boldsymbol{\alpha}$ and $\boldsymbol{\beta}$.  
The benefit of our method gradually increases as the network becomes deeper: the cost  ratio of \texttt{DD baseline} to ours increases modestly from approximately $20.8\times$ for three layers to slightly over $22\times$ for four and five layers. 
\subsubsection{\bf Impact of Solver Selection in CEDD-optimizer}
We then evaluate the impact of solver selection in the optimizer on overhead and quality, namely, the solver run time and the objective value. 
To this end, we compare the SLSQP solver with some alternatives: (1) a Trust-Region Constrained optimization (TRC) solver~\cite{trustregion,scipy_trust_constr}; (2) an Integer Differential Evolution (IDE) solver~\cite{IDE,IDE_link}; and (3) a Mixed-Integer Nonlinear Programming (MINLP) solver~\cite{gekko,MINLP}. 
Note that the SLSQP and TRC solvers are continuous, while the rest are discrete solvers. 
Therefore, continuous relaxation is applied to the former two.

The two leftmost graphs in Fig.~\ref{fig:extend3} compare the solvers in terms of run time and total cost, respectively. 
In this comparison, we set the edge count $N$ to 10. 
As shown in the graph~(a), the run time is only 0.006s for the SLSQP solver, whereas it takes 2.431s, 110.686s, and 2.623s for the TRC, IDE, and MINLP solvers, respectively. 
The graph~(b) reports the total cost for the aforementioned \texttt{Ideal} scenario. 
The total cost is broken down into (1) system cost and (2) optimizer (or solver) cost, which are normalized to the total cost of SLSQP.
The relative costs of TRC, IDE, and MINLP are $1.33$, $0.94$, and $0.91$, respectively. 
Consequently, the SLSQP solver outputs a solution comparable to those of the discrete solvers within \textit{orders of magnitude shorter run time}. 
Furthermore, these discrete solvers do not solve the problem in a practical amount of time when $N$ is scaled, due to the complex nature of the integer optimization. 
The SLSQP solver, therefore, is a favorable choice to balance solution quality, overhead, and scalability.


\subsubsection{\bf Order Selection in Test Accuracy Model}\label{sec:order-select}
Next, we validate the choice of polynomial degree and moment order in our test accuracy function formulated in Section~\ref{subsub:Statistical Accuracy Modeling}. 
To this end, we define the variables of skewness and kurtosis as follows: $
s_\alpha=\operatorname{skew}(\boldsymbol{\alpha}), 
s_\beta=\operatorname{skew}(\boldsymbol{\beta}), 
k_\alpha=\operatorname{kurt}(\boldsymbol{\alpha}), 
k_\beta=\operatorname{kurt}(\boldsymbol{\beta})
$. 
We compare the following 8 different models to represent the test accuracy function $Acc()$.  

\begin{itemize}

\item[\textbf{A.}] \textbf{First-order means (3 unknowns):} $a_0+a_1\mu_\alpha+a_2\mu_\beta$.

\item[\textbf{B.}] \textbf{First-order means and variances (5 unknowns):} $a_0+a_1\mu_\alpha+a_2\mu_\beta+a_6\sigma_\alpha^2+a_7\sigma_\beta^2$.

\item[\textbf{C.}] \textbf{Second-order means (6 unknowns):} $a_0+a_1\mu_\alpha+a_2\mu_\beta+a_3\mu_\alpha^2+a_4\mu_\alpha\mu_\beta
+a_5\mu_\beta^2$. 

\item[\textbf{D.}] \textbf{Second-order means and first-order variances (our choice, 8 unknowns):} $a_0+a_1\mu_\alpha+a_2\mu_\beta+a_3\mu_\alpha^2+a_4\mu_\alpha\mu_\beta+a_5\mu_\beta^2+a_6\sigma_\alpha^2+a_7\sigma_\beta^2$. 

\item[\textbf{E.}] \textbf{Second-order means and variances (11 unknowns):} $a_0+a_1\mu_\alpha+a_2\mu_\beta+a_3\mu_\alpha^2+a_4\mu_\alpha\mu_\beta+a_5\mu_\beta^2+a_6\sigma_\alpha^2+a_7\sigma_\beta^2+a_8\sigma_\alpha^4+a_9\sigma_\alpha^2\sigma_\beta^2+a_{10}\sigma_\beta^4$.

\item[\textbf{F.}] \textbf{Third-order means and second-order variances (15 unknowns):} $a_0+a_1\mu_\alpha+a_2\mu_\beta+a_3\mu_\alpha^2+a_4\mu_\alpha\mu_\beta+a_5\mu_\beta^2+a_6\sigma_\alpha^2+a_7\sigma_\beta^2+a_8\sigma_\alpha^4+a_9\sigma_\alpha^2\sigma_\beta^2+a_{10}\sigma_\beta^4+a_{11}\mu_\alpha^3+a_{12}\mu_\alpha^2\mu_\beta+a_{13}\mu_\alpha\mu_\beta^2+a_{14}\mu_\beta^3$. 


\item[\textbf{G.}] \textbf{Model D (our choice) plus first- and second-order skewness (13 unknowns):} 
$a_0+a_1\mu_\alpha+a_2\mu_\beta+a_3\mu_\alpha^2+a_4\mu_\alpha\mu_\beta+a_5\mu_\beta^2+a_6\sigma_\alpha^2+a_7\sigma_\beta^2+b_1s_\alpha+b_2s_\beta+b_3s_\alpha^2+b_4s_\alpha s_\beta+b_5s_\beta^2$. 


\item[\textbf{H.}] \textbf{Model G plus first- and second-order kurtosis (18 unknowns):} $a_0+a_1\mu_\alpha+a_2\mu_\beta+a_3\mu_\alpha^2+a_4\mu_\alpha\mu_\beta+a_5\mu_\beta^2+a_6\sigma_\alpha^2+a_7\sigma_\beta^2+b_1s_\alpha+b_2s_\beta+b_3s_\alpha^2+b_4s_\alpha s_\beta+b_5s_\beta^2+b_6k_\alpha+b_7k_\beta+b_8k_\alpha^2+b_9k_\alpha k_\beta+b_{10}k_\beta^2$. 


\end{itemize}

Fig.~\ref{fig:extend3}~(c)~/~(d) compares the above functions in absolute error / MAPE. 
The model D (our choice) achieves an absolute error of 0.73 and an MAPE of 1.31\%, closely matching those of the models E and F. 
This means that increasing the degree of polynomial of our model does not contribute significantly to error reduction. 
As for the order of statistical moment, scaling the order by adding more descriptive statistics gradually reduces the error --- see the errors of the models C, D, G, and H. 
However, a model with more unknowns requires more samples in the online calibration process presented in Section~\ref{subsection:three-step}. 
As an example, the model H requires at least 18 samples. 
Considering the cost-accuracy trade-off, we stopped at the second order moments by taking only means and variances. 

\section{Discussion}\label{discussion}\label{subsection:future-direction}

In this section, we discuss the impact of DD, as well as additional benefits beyond the results above, and discuss future challenges in its use.

\subsection{Privacy Protection} 
As DD strategically distills multiple images into a single image, it deliberately avoids revealing the original data during transmission.
Once sensitive information is removed during DD, it is extremely hard to restore it due to its lossy compression nature. 
We highlight a previous work by Dong~et~al.~\cite{dd-privacy} that theoretically proves this privacy benefit of DD. 
Other studies also report the privacy benefit~\cite{dd-privacy,dd-privacy1,dd-privacy2,dd-attacks1,dd-privacy4}. 
Although privacy protection is out of our scope, this preferred feature is another key motivation for and benefit of applying DD.

\subsection{Cross-Architecture Performance} 
The application of DD is agnostic to the target network architecture, which is later trained on the distilled dataset in the downstream training task. 
This is supported by previous studies demonstrating decent cross-architecture performance~\cite{dd-mtt,dd-att,dd-cafe,dd-kip,dd-dc,dd-dm}. 
This enables us to adopt a single, simple network architecture for our DD procedure, making it suitable for performing DD on the edges. 

\subsection{Coordination with Clock Frequency Scaling} 
Another benefit of our approach is the ability to incorporate hyperparameter tuning with power management mechanisms (e.g., clock frequency scaling) across edges. 
This extension is rather straightforward: simply find the best clock setup during the offline energy-coefficient calibration and apply it in the online DD phase. 
Consequently, this extension is fundamentally minor and orthogonal to our work, affecting only the energy model coefficients as functions of clock frequency.

\subsection{Application to Continual Learning} 
Continual Learning (CL) is a widely used learning paradigm in which models learn from a continuous stream of datasets. 
Existing studies~\cite{dd-continue1,dd-continue2,dd-continue3,dd-continue4,dd-continue5,dd-continue6} explore the use of DD in CL, which is also a potential use case of CEDD-Optimizer. 
In this use case, the calibration procedure for the accuracy model can be further optimized by leveraging CL's nature. 




\subsection{Application to Federated Learning} 
Another potential use case of  CEDD-optimizer is Federated Learning (FL), while this paper focuses on centralized learning.
Several studies employ DD in FL, replacing network parameters with decentralized updates~\cite {dd-fed0,dd-fed-early1,dd-fed2,dd-fed-3,dd-fed-4,dd-fed-early2}.
While the integration with FL is beyond the scope of this paper, our work also provides the foundation for this extension. 
Note that the concept of price-variation-aware hyperparameter optimization is valid for DD, regardless of whether the training host is centralized or distributed. However, the optimization strategy needs to be modified for FL due to differences in computation and communication patterns across edges. 



\subsection{Handling Non-IID Data}\label{sec:non-iid}
While CEDD-optimizer mainly targets IID and near-IID data, adapting to general non-IID data would be an immediate next step. 
First, DD is also valid for non-IID data as reported in various studies~\cite{dd-fed-heterougenious-data4,dd-fed2,dd-fed-4,dd-fed-heterougenious-data1,dd-fed-heterougenious-data2,dd-fed-heterougenious-data3}. 
In our CEDD-optimizer, our cost model is equally applicable to non-IID datasets, while the skewed amount of information across edges, induced by non-IID setting, can affect the test accuracy of downstream training tasks. 
To take such skewness into account in our test accuracy prediction, a potential solution is simply to use weighted basic statistics (e.g., weighted mean) as the model input ($\mu_{\alpha}$, $\mu_{\beta}$, $\sigma_{\alpha}$, and $\sigma_{\beta}$), while keeping the modeling structure unchanged. 
The key challenge here is quantifying the weights correctly to represent the amount of information each local dataset has. 
In this calculation, one can use sampled data, instead of using all the data, from the dataset, to limit the overhead.

\subsection{Temporal Variability of Electricity Prices}\label{sec:electricity}
Electricity prices can vary over time depending on the availability of renewable energy and the power demand. 
Although we did not take this effect into account in this paper, the variability offers another research opportunity. 
More specifically, optimizing the trigger timing of DD, using price forecast data, can be another optimization knob that operates on top of our present work.

\section{Conclusions}\label{conclusion}
This paper explored the use of DD for deep learning workflows deployed on edge devices or across the edge-to-cloud continuum, while accounting for the economic costs, including global network traffic and electricity for edge computing, which depend heavily on the geographical locations of the edge devices. 
To this end, we proposed a hyperparameter tuning framework, CEDD-optimizer, that leverages key hyperparameters of DD for geographically distributed edge devices to maximize the accuracy of downstream deep learning tasks while minimizing total economic cost. 
We formulated our tuning problem in a concrete mathematical form and provided predictive models that are incorporated into our solution workflow. 
We thoroughly evaluated our solution across various scenarios and parameter configurations, demonstrating the feasibility of DD in the context of deep learning workflow deployments on edge computing, where both service quality (in terms of deep learning task accuracy) and operating costs matter. 
We hope our study will inspire projects and solutions for the intrinsically coupled challenges between deep learning performance and economic operating cost.

\section*{Acknowledgments}
Performance results were partially obtained on systems in the test environment BEAST (Bavarian Energy Architecture \& Software Testbed) at the Leibniz Supercomputing Centre. This work was supported by the research project \textit{Optimierung von Gasturbinen mit Hilfe von Big Data} (AZ-1214-16), funded by the Bayerische Forschungsstiftung.

{
\bibliography{lib}
\bibliographystyle{IEEEtran}
}
\end{document}